\documentclass[aip,apl,amsmath,amssymb,reprint,superscriptaddress,preprintnumbers,showpacs,intlimits,shortbibliography]{revtex4-1}
\pdfoutput=1
\bibpunct{[}{]}{,}{n}{}{}

\usepackage[utf8]{inputenc}
\usepackage{bm,latexsym,mathrsfs,enumerate}
\usepackage[dvipsnames]{xcolor}
\usepackage{mathtools}
\usepackage{makecell}
\usepackage{multirow}
\usepackage[breaklinks=true,unicode=true,urlcolor = blue,colorlinks = true,citecolor = blue,linkcolor = blue]{hyperref}
\usepackage{graphicx}
\usepackage{ragged2e}
\usepackage[export]{adjustbox}
\renewcommand{\vec}[1]{\bm{#1}}
\usepackage[final]{pdfpages}
\usepackage{tikz}
\makeatletter
\AtBeginDocument{\let\LS@rot\@undefined}
\makeatother

\begin{document}
\title{Domain wall diode based on functionally graded Dzyaloshinskii--Moriya interaction}

\author{Kostiantyn~V.~Yershov}
\email[Corresponding author: ]{yershov@bitp.kiev.ua}
\affiliation{Bogolyubov Institute for Theoretical Physics of National Academy of Sciences of Ukraine, 03143 Kyiv, Ukraine\looseness=-1}
\affiliation{Leibniz-Institut f\"ur Festk\"orper- und Werkstoffforschung, IFW Dresden, D-01171 Dresden, Germany\looseness=-1}

\author{Volodymyr~P.~Kravchuk}
\email[ ]{vkravchuk@bitp.kiev.ua}
\affiliation{Bogolyubov Institute for Theoretical Physics of National Academy of Sciences of Ukraine, 03143 Kyiv, Ukraine\looseness=-1}
\affiliation{Institut f\"ur Theoretische Festk\"orperphysik, Karlsruher Institut f\"ur Technologie, D-76131 Karlsruhe, Germany\looseness=-1}

\author{Denis~D.~Sheka}
\email[ ]{sheka@knu.ua}
\affiliation{Taras Shevchenko National University of Kyiv, 01601 Kyiv, Ukraine\looseness=-1}

\author{Jeroen van den Brink}
\email[ ]{j.van.den.brink@ifw-dresden.de}
\affiliation{Leibniz-Institut f\"ur Festk\"orper- und Werkstoffforschung, IFW Dresden, D-01171 Dresden, Germany\looseness=-1}
\affiliation{Institute for Theoretical Physics, TU Dresden, 01069 Dresden, Germany\looseness=-1}

\author{Avadh Saxena}
\email[ ]{avadh@lanl.gov}
\affiliation{Theoretical Division, Los Alamos National Laboratory, Los Alamos, NM 87545, USA\looseness=-1}


%
%
%
%
\begin{abstract}

We present a general approach for studying the dynamics of domain walls in biaxial ferromagnetic stripes with functionally graded Dzyaloshinskii--Moriya interaction~(DMI).  By engineering the spatial profile of the DMI parameter we propose the concept of a diode, which implements filtering of domain walls of certain topological charge and helicity. We base our study on phenomenological Landau--Lifshitz--Gilbert equations with additional Zhang--Li spin-transfer terms using a collective variable approach. In the effective equations of motion the gradients of DMI play the role of a driving force which competes with current driving. All analytical predictions are confirmed by numerical simulations.

\end{abstract}

\maketitle

%
%

\textit{Introduction.} Topological spin textures have ignited a growing interest in spintronics due to their rich phenomenology as well as novel potential applications. Their nanoscale size and topologically-protected stability make them attractive candidates for information carriers in high-density data-storage technologies. For example, domain walls~(DW) and skyrmions in magnetic nanostripes are proposed as key elements of nonvolatile magnetic logic~\cite{Allwood05,Krause16} and  memory~\cite{Xu08a,Parkin08,Woo16} devices. Magnetic systems with functionally graded internal material parameters are particularly promising for this use. Recently, spatial engineering of the anisotropy~\cite{Franken12a,Zhang16b,Sanchez18,Ang19} and Dzyloshinskii--Moriya interaction~(DMI)~\cite{Diaz16,Hong17,Zhou19,Menezes19,Toscano19} profiles have been suggested as an alternative way of DWs and skyrmions guidance and manipulation. The interest in these results is stimulated by the fact that systems with functionally-graded material parameters can be fabricated experimentally. For instance, it has been shown that the Bloch DMI can be controlled by the chemical composition~\cite{Grigoriev09,Grigoriev10,Shibata13,Grigoriev13,Morikawa13,Siegfried15,Koretsune15} in chiral ferromagnetic materials. On the other hand, the N{\'e}el DMI in multilayer thin films can be tuned by the thickness of the Pt layer~\cite{Ma16a,Tacchi17}, the thickness of the ferromagnetic layer~\cite{Nembach15,Belmeguenai15,Stashkevich15,Lee15a,Kopte17}, electric field~\cite{Srivastava18,Yang18b}, and by ion irradiation~\cite{Balk17}.

In this paper, we study the current induced dynamics of DWs in a chiral ferromagnetic film with functionally graded DMI. We show that the gradient of DMI parameter results in the driving force for DWs similarly to the curvature gradient in wires~\cite{Yershov15b} and  stripes~\cite{Yershov18a}. Considering the coordinate dependent DMI parameter we propose a general approach valid for an arbitrary profile of the DMI parameter distribution. We also show how the competition of DMI-induced driving force and current pumping can be potentially used in applications.

%
%

\textit{Model and general results.} We consider a thin and narrow ferromagnetic stripe whose thickness and width are small enough to ensure the one-dimensional character of changes in the magnetization, and the stripe length significantly exceeds the lateral dimensions. Thus, the magnetization is described by a continuous and normalized function $\vec{m}=\vec{M}/M_s=\vec{m}(x,t)$, where $M_s$ is the saturation magnetization, $x$-axis is orientated along the stripe, and $t$ denotes time. The magnetization dynamics is described by the Landau--Lifshitz--Gilbert equation with additional Zhang--Li spin-torque terms \cite{Bazaliy98,Zhang04,Thiaville05}

\begin{figure*}[t]
	\includegraphics[width=\textwidth]{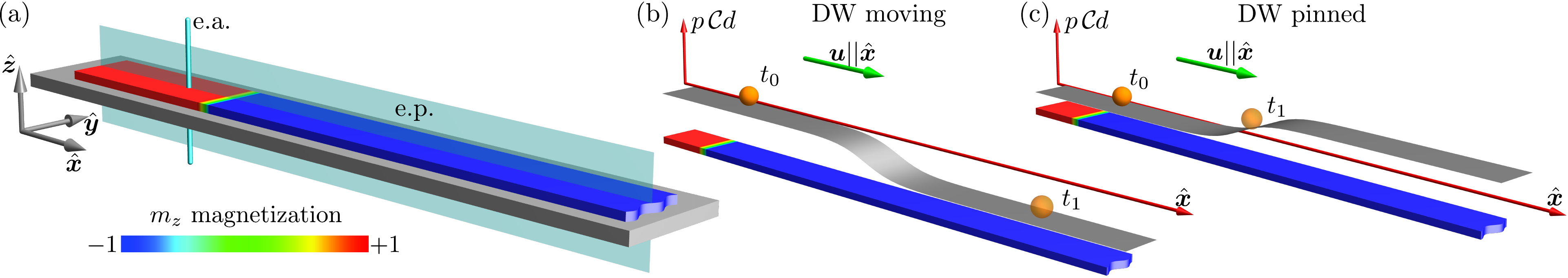}
	\caption{\label{fig:scheme}%
		(Color online) (a) DW in a biaxial stripe, e.a. and e.p. denotes ``easy axis" and ``easy plane", respectively. (b),(c) Possible regimes of DW motion: (b) DW is freely moving, while in (c) DW is pined. The gray surface in (b) and (c) demonstrates the profile of the DW DMI energy $\mathcal{E}_\textsc{dm}\propto p\,\mathcal{C}d[q(t)]$ as a function of DW position~$q(t)$~(orange balls correspond to the DW position at different moments of time $t_1>t_0$), $p$ and $\mathcal{C}$ are the DW topological charge and helicity, respectively, and $d$ denotes the dimensionless DMI parameter.}
\end{figure*}

\begin{equation}\label{eq:LLG_ZhLi}
\begin{split}
\frac{\partial\vec{m}}{\partial t}=&\;\omega_0\,\vec{m}\times\frac{\delta\mathcal{E}}{\delta\vec{m}} + \alpha\,\vec{m}\times\frac{\partial\vec{m}}{\partial t}\\
+&\,\vec{m} \times \left[\vec{m}\times\left(\vec{u}\cdot\nabla\right) \vec{m}\right] + \beta\,\vec{m} \times \left(\vec{u}\cdot\nabla\right)\vec{m},
\end{split}
\end{equation} 
where $\omega_0=\gamma_0 K/M_s$  determines the characteristic timescale of the system with $\gamma_0$ being the gyromagnetic ratio. Here $\mathcal{E}=E/K$ is normalized total energy of the system, where $K>0$ is the easy-axis anisotropy constant, see Fig.~\ref{fig:scheme}(a). The driving strength is represented by the quantity $\vec{u}=\vec{j}P\mu_\textsc{b}/(|e|M_s)$ which is close to average electron drift velocity in the presence of a current of density $\vec{j}||\hat{\vec{x}}$. Here $P$ is the rate of spin polarization, $\mu_\textsc{b}$ is Bohr magneton, and $e$ is electron charge. Constants $\alpha$ and $\beta$ denote Gilbert damping and the nonadiabatic spin-transfer parameter, respectively.

To write the energy functional $\mathcal{E}$ we consider a simple model, which takes into account only three contributions to the total magnetic energy:
\begin{equation}\label{eq:energy_model}
\mathcal{E}=\mathcal{S}\int_{-\infty}^{+\infty}\left[\ell^2\mathscr{E}_\text{ex}+\mathscr{E}_\text{an}+\frac{D(x)}{K}\mathscr{E}_\textsc{dm}\right]\textrm{d}x,
\end{equation}
where $\mathcal{S}$ is the stripe cross-section area. The first term in~\eqref{eq:energy_model} is the exchange energy density $\mathscr{E}_\text{ex}=\sum_{i=x,y,z}\left(\partial_i\vec{m}\right)^2$. The competition between exchange and anisotropy results in the magnetic length $\ell=\sqrt{A/K}$, which determines a length scale of the system, here $A$ is the exchange constant and K is the easy-axis anisotropy constant. The second term in~\eqref{eq:energy_model} corresponds to the biaxial anisotropy contribution $\mathscr{E}_\text{an}=1-m_z^2+\varepsilon\, m_y^2$, where $\varepsilon = K_p / K$ with $K_p>0$ being the easy-plane anisotropy coefficient. The easy-axis is perpendicular to the stripe plane~($xy$-plane), while easy-plane coincides with the $xz$-plane, see Fig.~\ref{fig:scheme}. Such kind of anisotropy is effectively induced by the magnetostatic interaction in the thin stripes with the perpendicular easy-axis magnetocrystalline anisotropy~\cite{Aharoni98,Porter04,Hillebrands06}. For thin and narrow stripes the approximation of the shape anisotropy is used also for inhomogeneous magnetization states, e.g.~DWs~\cite{Mougin07,Yershov18a}.  The last term in~\eqref{eq:energy_model} corresponds to the N{\'e}el DMI $\mathscr{E}_\textsc{dmi}^\textsc{n}=m_z\vec{\nabla}\cdot\vec{m}-\vec{m}\cdot\vec{\nabla}m_z$ and $D(x)$ describes the spatial profile of DMI strength. This type of DMI is taken in the form typical for ultrathin films~\cite{Bogdanov01,Thiaville12}, bilayers~\cite{Yang15} or materials belonging to the $C_{nv}$ crystallographic group~\cite{Leonov16}.

Since $|\vec{m}|=1$, it is convenient to proceed to the angular representation $\vec{m}=\vec{e}_x\sin\theta\cos\phi+\vec{e}_y\sin\theta\sin\phi+\vec{e}_z\cos\theta$, where $\theta=\theta(x)$ and $\phi=\phi(x)$ are magnetic angles. In terms of angular parametrization, the energy density in~\eqref{eq:energy_model} has the following form
\begin{equation}\label{eq:energy_density}
	\begin{split}
		\mathscr{E}=\left(\theta'\right)^2+\left(\phi'\right)^2&\sin^2\theta+\sin^2\theta\left(1+\varepsilon\sin^2\phi\right)\\
		+d&\left(\theta'\cos\phi-\frac{\phi'}{2}\sin2\theta\sin\phi\right).
	\end{split}
\end{equation}
Here and below prime denotes the derivative with respect to the dimensionless coordinate $\xi=x/\ell$, and $d(\xi)=D/\sqrt{AK}$ is a dimensionless DMI parameter.

Let us first analyze static magnetization distribution determined by the minimum of the energy~\eqref{eq:energy_model}. Minimization of~\eqref{eq:energy_model} with associated energy density~\eqref{eq:energy_density} with respect to $\phi$ results in a solution $\cos\phi=\cos\phi_0=\mathcal{C}=\pm1$~\footnote[1]{Here we consider the spatial distribution of DMI strength without the change of sign, i.e. $d\geq 0$.}. This means that vectors $\vec{m}$ lie within the $xz$-plane. The corresponding function $\theta$ is determined by a driven pendulum equation~\footnote[2]{See Supplemental Material at \texttt{Link provided by the publisher} for details of analytical calculations and movies, which includes Refs.~\cite{Bazaliy98,Zhang04,Thiaville05,Doering48,Hillebrands06,Mougin07,Yershov16}.}
\begin{equation}\label{eq:theta_def}
	\theta''-\sin \theta\cos\theta=-\frac{d'}{2}\cos\phi_0.
\end{equation}
Equation~\eqref{eq:theta_def} is analogous to one, which determines the DW structure in flat curved wires~\cite{Yershov15b} and stripes~\cite{Yershov18a}, where curvature results in a coordinate-dependent effective DMI.

For the case $d'\equiv 0$ Eq.~\eqref{eq:theta_def} has a well known DW solution $\cos\theta=-p\tanh\left[\left(\xi-q\right)/\Delta\right]$, where $q$ is the DW position, $p=\pm1$ being the topological charge~($p=+1$: kink, $p=-1$: antikink), and  $\Delta$ is the DW width. Here~$q$ and $\Delta$ are dimensionless quantities measured in units of~$\ell$. For the case $d'\not=0$ an additional driving force  appears similarly to the case discussed in Refs.~\onlinecite{Yershov15b} and \onlinecite{Yershov18a}. In the following, we consider a case of the spatial distribution of the DMI strength with $d'\left(\pm\infty\right)=0$, which allows the boundary conditions $\cos\theta(\pm\infty)=\mp p$. We restrict ourselves to the case $d< 4/\pi$ and consider $d'$ as a small perturbation which does not modify significantly the profile of the DW and its width $\Delta$. Therefore, to analyze the DW properties we use the collective variable approach based on the $q$--$\Phi$ model~\cite{Slonczewski72,Malozemoff79}
\begin{equation}\label{eq:q_phi_model}
\cos\theta=-p\tanh\frac{\xi-q(t)}{\Delta},\quad \phi=\Phi(t).
\end{equation}
Here, $\{q,\Phi\}$ are time-dependent conjugated collective variables, which determine the DW position and phase, respectively. The DW width $\Delta$ is assumed to be  a slaved variable~\cite{Hillebrands06}, i.e. $\Delta(t) = \Delta\left[\Phi\left(t\right),q\left(t\right)\right]$.

\begin{figure*}[t]
	\includegraphics[width=\textwidth]{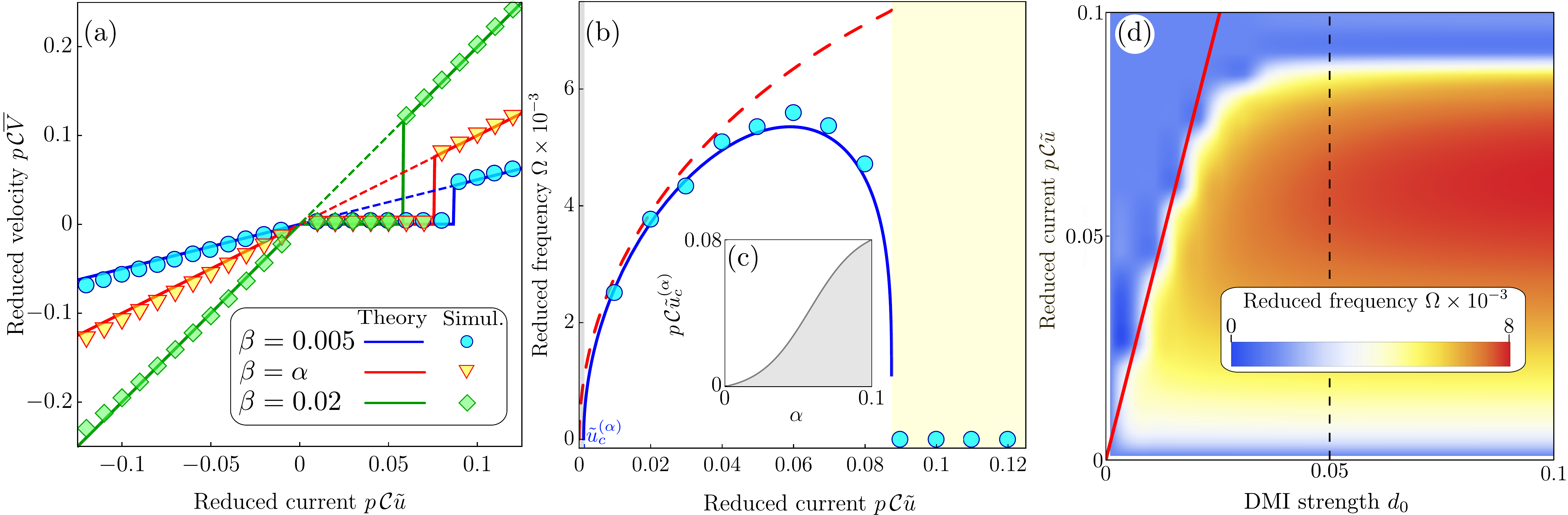}
	\caption{\label{fig:velocity_freq}%
		(Color online) (a) Averaged DW velocity as a function of applied current. Solid lines correspond to solutions of the collective variables equations~\eqref{eq:q_phi_current}. Dashed lines correspond to a DW velocity~\eqref{eq:q_phi_travel}. (b)~and~(c)~Eigenfrequency of DW oscillations in the vicinity of the pinning position. Solid and dashed lines are plotted with prediction (S11)~\cite{Note2} and its approximation for zero damping and small currents~\eqref{eq:omega_eta}, respectively; filled gray area in (b) and (c) corresponds to the overdamped regime when DW is pinned without oscillations with $\Omega=0$; filled yellow area corresponds to the currents with  $\tilde{u}>\tilde{u}_c$. (d) The eigenfrequency of DW oscillations in the vicinity of the pinning position in terms of density plot. Solid line corresponds to the depinning current $\tilde{u}_c=\pi d_0/(4\beta w)$. Symbols in (a) and (b) correspond to the numerical simulations with $d_0=0.05$, see vertical dashed line in (d). All other parameters of simulations: $\varepsilon=0.1$, $\alpha=0.01$, $p=+1$, $\tilde{u}>0$, $w=10$, and stripe width $2\ell$. In (b)-(d) we have $\beta=0.02$.}
\end{figure*}

Substituting the Ansatz~\eqref{eq:q_phi_model} into~\eqref{eq:energy_density} and performing integration over the $\xi$ coordinate, we obtain the energy of a DW in the stripe in the form
\begin{equation}\label{eq:energy_DW}
\frac{\mathcal{E}}{2\mathcal{S}\ell}=\frac{1}{\Delta}+\Delta\left(1+\varepsilon\sin^2\Phi\right)+p\frac{d(q)}{2}\pi\cos\Phi,
\end{equation}
where the condition $\Delta\,d'\ll1$ was imposed when integrating~\eqref{eq:energy_model}. The structure of the energy~\eqref{eq:energy_DW} has similar form as DW energy in a curved biaxial stripe~\cite{Yershov18a}. The first two terms on the right hand side in~\eqref{eq:energy_DW} determine the competition of the isotropic exchange and anisotropy contributions, while the third term originates from the DMI and demonstrates the coupling between the DMI strength $d$, DW topological charge $p$, and helicity $\mathcal{C}$, i.e. DMI energy is minimized when $\mathcal{C}=-\text{sgn}\left(pd\right)$. In the following, a DW which corresponds to the global minimum of the energy \eqref{eq:energy_DW} in the parametric space $\{q,\Phi,\Delta\}$ is called \textit{favorable}. Under the condition $|d|/\varepsilon<4/\pi$, energy \eqref{eq:energy_DW} also has a local minimum, which corresponds to a DM with the opposite helicity $-\mathcal{C}$. In the following this DW is called \textit{unfavorable}.

In terms of collective variables, the equations of motion~\eqref{eq:LLG_ZhLi} take the form~\cite{Note2}
\begin{equation}\label{eq:q_phi_current}
\begin{split}
\frac{\alpha}{\Delta}\dot{q}+p\dot{\Phi}&=-p\frac{\pi}{2}\frac{\partial d(q)}{\partial q}\cos\Phi+\frac{\beta}{\Delta}\tilde{u},\\
p\dot{q}-\alpha\Delta\dot{\Phi}&=-p\frac{\pi}{2}d(q)\sin\Phi+\varepsilon\Delta\sin2\Phi+p\tilde{u},
\end{split}
\end{equation}
where overdots indicate the derivative with respect to the dimensionless time $\tau=t\omega_0$ and $\tilde{u}=u/\left(\omega_0\ell\right)$ is a dimensionless current. The DW width is $\Delta(\tau)=\Delta[\Phi(\tau)]=1/\sqrt{1+\varepsilon\sin^2\Phi}$. The behavior of the DW width is discussed in detail in the Supplemental Material~\cite{Note2}.

For the case $\tilde{u}=0$ Eqs.~\eqref{eq:q_phi_current} coincide with equations of motion for DW in a curved biaxial stripe presented in Ref.~\onlinecite{Yershov18a} with the curvature gradient replaced by the gradient of DMI strength. The DMI induced driving force can suppress the action of the pumping by the spin-polarized current or can reinforce it, see Fig.~\ref{fig:velocity_freq}. In other words, the unfavorable (favorable) DW has to overcome the energetic barrier, when it enters the region with larger (smaller) $|d|$. If the applied current is small enough, then such a DW is pinned in position $\left[d(q_\text{pin})-d(q_0)\right]/\left[q_\text{pin}-q_0\right]=2p\,\mathcal{C}\beta\tilde{u}/\pi$. For small currents the phase of the pinned DW does not deviate significantly from its equilibrium value:  $\Phi_\text{pin}\approx \{0,\pi\}$~\cite{Note2}. The average DW velocity as a function of current for the DMI profile
\begin{equation}
d=\frac{d_0}{2}\left[\tanh\left(\frac{\xi}{w}\right)+1\right]
\end{equation}
with amplitude $d_0$ and width $w$ is presented in Fig.~\ref{fig:velocity_freq}(a), also see Supplemental movies~\cite{Note2} for the corresponding DW dynamics. Zero averaged velocity~\footnote[3]{The velocity of DW is calculated as $\overline{V}=T^{-1}\int_{0}^{T}\dot{q}(\tau)\mathrm{d}\tau$, where $\dot{q}(\tau)$ is extracted from numerical simulations and $T\gg 1$ is a time simulation.} corresponds to the case of pinning.

The behavior of DW velocity presented in Fig.~\ref{fig:velocity_freq}(a) demonstrates that functionally graded DMI allows to filter DWs of certain type, i.e. we have built a DW diode in a planar nanostripe. Depinning of DWs takes place when current $\tilde{u}$ exceeds some critical value~$\tilde{u}_c\approx\pi\,\text{max}_{q}\left[d'\left(q\right)\right]/\left(2\beta\right)$, see Fig.~\ref{fig:velocity_freq}(d). 

Next, we study linear dynamics of the DW in the vicinity of the DW pinning position. With this purpose we introduce small deviations as $q(\tau) = q_\text{pin} + \tilde{q}(\tau)$ and~$\Phi\left(\tau\right) = \Phi_\text{pin}+\tilde{\Phi}(\tau)$. The equations of motion~\eqref{eq:q_phi_current} linearized with respect to the deviations read~as
\begin{equation}\label{eq:q_phi_linearized}
\left(1+\alpha^2\right)\left\|\begin{matrix}
\dot{\tilde{q}}\\
\dot{\tilde{\Phi}}
\end{matrix}\right\|=\hat{\mathcal{M}}\left\|\begin{matrix}
\tilde{q}\\
\tilde{\Phi}
\end{matrix}\right\|,
\end{equation}
where matrix $\hat{\mathcal{M}}=\hat{\mathcal{M}}(\tilde{u})$ depends on current~\cite{Note2}. For the case of low damping and low current the solution of~\eqref{eq:q_phi_linearized} results in  decaying oscillations with frequency 
\begin{equation}\label{eq:omega_eta}
\Omega\approx \frac{\pi}{2}\sqrt{d''(q_\text{pin})\left[p\,\mathcal{C}\varepsilon\frac{4}{\pi}-d\left(q_\text{pin}\right)\right]}.
\end{equation}
The frequency $\Omega$ as a function of applied current is plotted in Figs.~\ref{fig:velocity_freq}(b) and \ref{fig:velocity_freq}(d). Depinning of the DW takes place for cases: (i) $d''(q_\text{pin})= 0$~[see Fig.~\ref{fig:velocity_freq}(b)] or (ii) $|d(q_\text{pin})|/\varepsilon>4/\pi$. In the latter case, the unfavorable DW experiences the phase flip $\Phi \to \Phi+\pi$ and transforms to the favorable DW. The limiting case of $d=\text{const}$ does not produce any pinning due to the zero gradient of the DMI strength.

\begin{figure}[t]
	\includegraphics[width=\columnwidth]{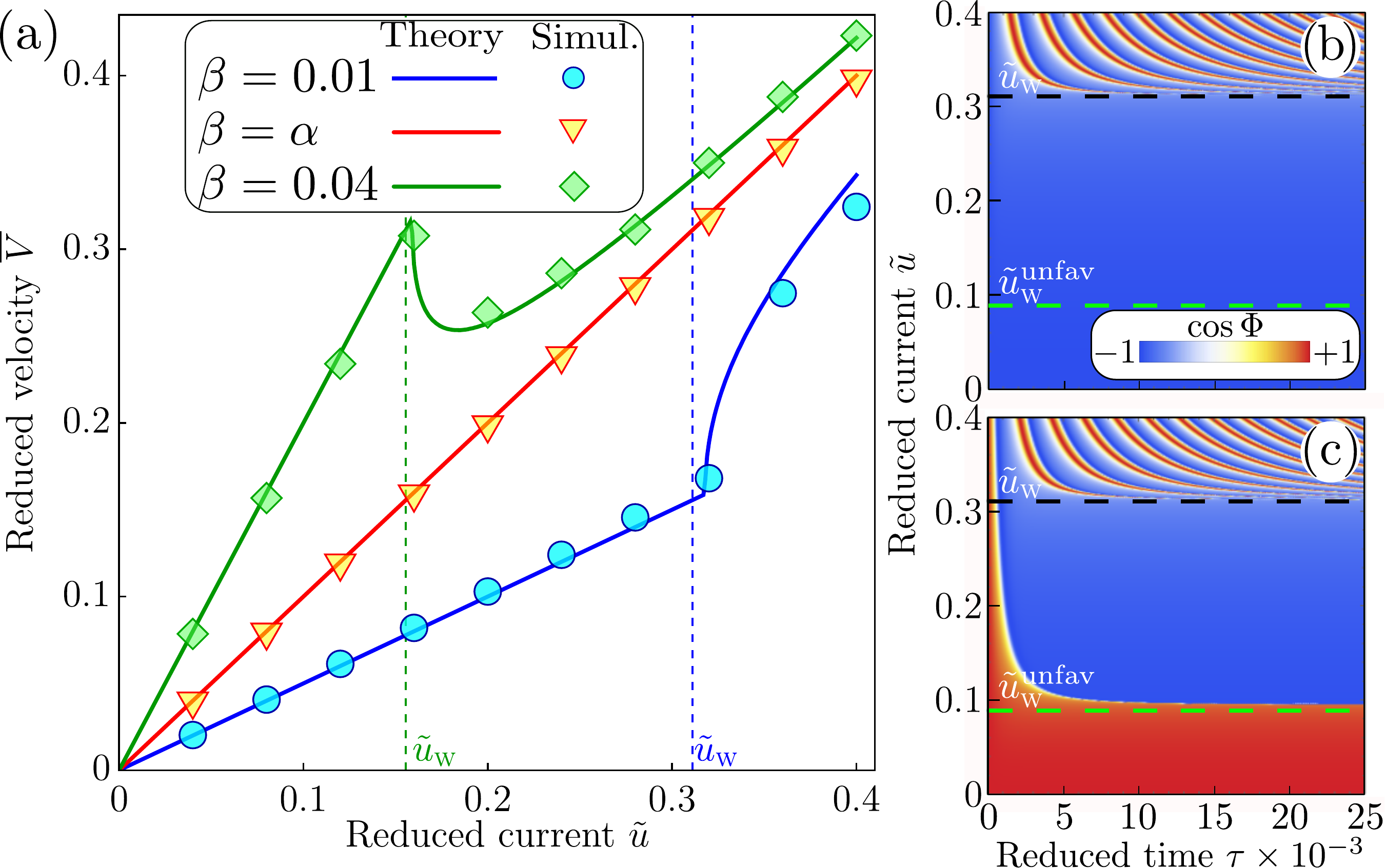}
	\caption{\label{fig:walker_current}%
		(Color online) (a) Averaged velocity of a favorable DW as a function of the applied current. Lines correspond to solutions of the collective variables equations~\eqref{eq:q_phi_current} with $d=d_0=\text{const}$, and symbols show the results of numerical simulations. Dashed lines correspond to the Walker current~\eqref{eq:walker_current}. (b) and (c) Show evolution of $\cos\Phi$ for DW for different values of current in terms of density plot for favorable and unfavorable DWs, respectively [data obtained from solutions of the collective variables equations~\eqref{eq:q_phi_current} with $d=d_0=\text{const}$ and $\beta=0.01$]. In all simulations we use $\alpha=0.02$, $d_0=0.05$, $p=+1$, $\varepsilon=0.1$, and stripes of width $2\ell$.}
\end{figure}

\textit{Case with zero gradient of the DMI strength.} Now we will consider the case with $d=d_0=\text{const}$. For this case  Eqs.~\eqref{eq:q_phi_current} have a solution for the traveling wave regime with $q=V\tau$ and $\Phi=\text{const}$~(the case $d_0=0$ discussed in Ref.~\onlinecite{Thiaville05}). The corresponding DW velocity and phase~(in the small current approximation) are
\begin{equation}\label{eq:q_phi_travel}
V=\frac{\beta}{\alpha}\tilde{u},\quad \Phi\approx\Phi_0+ \frac{2p}{4\varepsilon- p\, \mathcal{C}\pi d_0}\left(\frac{\beta-\alpha}{\alpha}\right)\tilde{u}.
\end{equation}
It is necessary to mention that DW velocity~\eqref{eq:q_phi_travel} is independent of the DMI parameter and coincides with DW velocity reported in Ref.~\onlinecite{Thiaville05}, see Fig.~\ref{fig:velocity_freq}(a). The traveling-wave solution~\eqref{eq:q_phi_travel} exists for the currents $\tilde{u} < |\tilde{u}_\textsc{w}|$, where
\begin{equation}\label{eq:walker_current}
\tilde{u}_\textsc{w}\approx\tilde{u}_\textsc{w}^0\pm\frac{\alpha}{|\alpha-\beta|}\left(\frac{\pi}{2\sqrt{2}}\right)d_0,\quad \tilde{u}_\textsc{w}^0 = \frac{\alpha}{|\alpha-\beta|}\varepsilon.
\end{equation}
Here ``$+$" sign corresponds to the \textit{favorable} DW, while ``$-$" sign corresponds to the \textit{unfavorable} DW~\cite{Note2}, and $\tilde{u}_\textsc{w}^0$ is the Walker current for DMI-free biaxial ferromagnetic system~\cite{Thiaville05}. Estimation \eqref{eq:walker_current} is obtained under the assumption $d_0\ll\varepsilon\ll1$. From \eqref{eq:walker_current} it follows that the DMI results in the shift of the Walker current in the biaxial stripe similarly to the case of field-driven DW~\cite{Kravchuk14}. The value of the Walker current for \textit{unfavorable} DW is smaller as compared to the \textit{favorable} one and it defines the current of the DW phase flip,  i.e. for $\tilde{u}_\textsc{w}^\text{unfav}<\tilde{u}<\tilde{u}_\textsc{w}$ we have traveling wave motion with a single flip of  the phase with $\Phi\to\Phi+\pi$, see Fig.~\ref{fig:walker_current}(b) and~\ref{fig:walker_current}(c). One should note that for the case $\beta=\alpha$ both DWs move in a traveling-wave regime without any flip of the phase $\Phi$, i.e. $\tilde{u}_\textsc{w}(\beta=\alpha)\to\infty$. The average DW velocity as a function of current is presented in Fig.~\ref{fig:walker_current}.

Let us estimate the effective mass of the DW~\cite{Doering48}. To this end, we consider a no driving case ($u = 0$) with vanishing damping. In this case a small deviation $\varphi= \Phi - \Phi_0$ of the DW phase from its equilibrium value results in the traveling-wave DW motion with the velocity $V\approx\left[2p\varepsilon-\mathcal{C}\pi d_0/2\right]\varphi$. Using the latter relation and energy expression~\eqref{eq:energy_DW} one can estimate the energy of the moving DW as $\mathcal{E}/\left(\mathcal{S}\ell\right) \approx \mathcal{E}_0/\left(\mathcal{S}\ell\right) + \mu V^2/2$, where $\mathcal{E}_0$ is the energy of a stationary DW and the quantity
\begin{equation}\label{eq:dw_mass}
\mu=\frac{4}{4\varepsilon-p\,\mathcal{C}\pi d_0}
\end{equation}
can be interpreted as the effective mass of the DW. For the DMI-free case~($d_0=0$) the effective mass~\eqref{eq:dw_mass} coincides with the D{\"o}ring mass~\cite{Doering48}.
The DMI results in the modification of the DW mass, i.e. for \textit{favorable}~(\textit{unfavorable}) DW the mass $\mu$ decreases~(increases) with the DMI strength, respectively.
 
\textit{Conclusions.} We have demonstrated that presence of the biaxial anisotropy~($\varepsilon>0$) and DMI~($|d|/\varepsilon <4/\pi$) allows the existence of DWs with different combinations of topological charge and helicity~($p\,\mathcal{C}=\pm1$). One of these DWs becomes energetically \textit{favorable}, i.e. it minimizes the DMI energy $\mathcal{E}_\textsc{dm}\propto p\,\mathcal{C}d$. By engineering the profile of DMI strength the \textit{favorable} DW will move to the area with a bigger DMI strength and will be pinned for opposite direction, while behavior for the \textit{unfavorable} DW is vice versa. This effect can be used for the fabrication of a DW diode, traps, and ratchets in a planar stripe. These can be potentially used for the development of logic devices. We show that the gradient of functionally graded DMI in magnetic systems results in the driving force for DWs. The competition between  the DMI-driving force and pumping by the current determines the behavior of DW dynamics. The intrinsic DMI results in the shift of the Walker current~\eqref{eq:walker_current}, which allows to increase the maximal velocity of traveling-wave motion for the DW; DMI modifies the DW D{\"o}ring mass~\eqref{eq:dw_mass}. This shows that functionally graded materials open new possibilities in the manipulation of DWs. We expect that the similar pinning effects can appear for magnetic skyrmions as well.

\textit{Acknowledgments.} We thank  U. Nitzsche for technical support. K.V.Y. acknowledges financial support from UKRATOP-project~(funded by BMBF under reference 01DK18002). JvdB acknowledges support from the German Research Foundation (Deutsche Forschungsgemeinschaft, DFG) via SFB1143 project A5 and through the W\"urzburg-Dresden Cluster of Excellence on Complexity and Topology in Quantum Matter - ct.qmat (EXC 2147, project id 39085490). In part, this work was supported by the Program of Fundamental Research of the Department of Physics and Astronomy of the National Academy of Sciences of Ukraine~(Project No.~0116U003192), by the Alexander von Humboldt Foundation (Research Group Linkage Programme), by Taras Shevchenko National University of Kyiv (Project No. 19BF052-01), and by the US Department of Energy.


\begin{thebibliography}{52}%
	\makeatletter
	\providecommand \@ifxundefined [1]{%
		\@ifx{#1\undefined}
	}%
	\providecommand \@ifnum [1]{%
		\ifnum #1\expandafter \@firstoftwo
		\else \expandafter \@secondoftwo
		\fi
	}%
	\providecommand \@ifx [1]{%
		\ifx #1\expandafter \@firstoftwo
		\else \expandafter \@secondoftwo
		\fi
	}%
	\providecommand \natexlab [1]{#1}%
	\providecommand \enquote  [1]{``#1''}%
	\providecommand \bibnamefont  [1]{#1}%
	\providecommand \bibfnamefont [1]{#1}%
	\providecommand \citenamefont [1]{#1}%
	\providecommand \href@noop [0]{\@secondoftwo}%
	\providecommand \href [0]{\begingroup \@sanitize@url \@href}%
	\providecommand \@href[1]{\@@startlink{#1}\@@href}%
	\providecommand \@@href[1]{\endgroup#1\@@endlink}%
	\providecommand \@sanitize@url [0]{\catcode `\\12\catcode `\$12\catcode
		`\&12\catcode `\#12\catcode `\^12\catcode `\_12\catcode `\%12\relax}%
	\providecommand \@@startlink[1]{}%
	\providecommand \@@endlink[0]{}%
	\providecommand \url  [0]{\begingroup\@sanitize@url \@url }%
	\providecommand \@url [1]{\endgroup\@href {#1}{\urlprefix }}%
	\providecommand \urlprefix  [0]{URL }%
	\providecommand \Eprint [0]{\href }%
	\providecommand \doibase [0]{http://dx.doi.org/}%
	\providecommand \selectlanguage [0]{\@gobble}%
	\providecommand \bibinfo  [0]{\@secondoftwo}%
	\providecommand \bibfield  [0]{\@secondoftwo}%
	\providecommand \translation [1]{[#1]}%
	\providecommand \BibitemOpen [0]{}%
	\providecommand \bibitemStop [0]{}%
	\providecommand \bibitemNoStop [0]{.\EOS\space}%
	\providecommand \EOS [0]{\spacefactor3000\relax}%
	\providecommand \BibitemShut  [1]{\csname bibitem#1\endcsname}%
	\let\auto@bib@innerbib\@empty
	\bibitem [{\citenamefont {Allwood}\ \emph {et~al.}(2005)\citenamefont
		{Allwood}, \citenamefont {Xiong}, \citenamefont {Faulkner}, \citenamefont
		{Atkinson}, \citenamefont {Petit},\ and\ \citenamefont
		{Cowburn}}]{Allwood05}%
	\BibitemOpen
	\bibfield  {author} {\bibinfo {author} {\bibfnamefont {D.~A.}\ \bibnamefont
			{Allwood}}, \bibinfo {author} {\bibfnamefont {G.}~\bibnamefont {Xiong}},
		\bibinfo {author} {\bibfnamefont {C.~C.}\ \bibnamefont {Faulkner}}, \bibinfo
		{author} {\bibfnamefont {D.}~\bibnamefont {Atkinson}}, \bibinfo {author}
		{\bibfnamefont {D.}~\bibnamefont {Petit}}, \ and\ \bibinfo {author}
		{\bibfnamefont {R.~P.}\ \bibnamefont {Cowburn}},\ }\href {\doibase
		10.1126/science.1108813} {\bibfield  {journal} {\bibinfo  {journal}
			{Science}\ }\textbf {\bibinfo {volume} {309}},\ \bibinfo {pages} {1688}
		(\bibinfo {year} {2005})}\BibitemShut {NoStop}%
	\bibitem [{\citenamefont {Krause}\ and\ \citenamefont
		{Wiesendanger}(2016)}]{Krause16}%
	\BibitemOpen
	\bibfield  {author} {\bibinfo {author} {\bibfnamefont {S.}~\bibnamefont
			{Krause}}\ and\ \bibinfo {author} {\bibfnamefont {R.}~\bibnamefont
			{Wiesendanger}},\ }\href {\doibase 10.1038/nmat4615} {\bibfield  {journal}
		{\bibinfo  {journal} {Nature Materials}\ }\textbf {\bibinfo {volume} {15}},\
		\bibinfo {pages} {493} (\bibinfo {year} {2016})}\BibitemShut {NoStop}%
	\bibitem [{\citenamefont {Xu}\ \emph {et~al.}(2008)\citenamefont {Xu},
		\citenamefont {Xia}, \citenamefont {Gu}, \citenamefont {Tang}, \citenamefont
		{Yang},\ and\ \citenamefont {Li}}]{Xu08a}%
	\BibitemOpen
	\bibfield  {author} {\bibinfo {author} {\bibfnamefont {P.}~\bibnamefont
			{Xu}}, \bibinfo {author} {\bibfnamefont {K.}~\bibnamefont {Xia}}, \bibinfo
		{author} {\bibfnamefont {C.}~\bibnamefont {Gu}}, \bibinfo {author}
		{\bibfnamefont {L.}~\bibnamefont {Tang}}, \bibinfo {author} {\bibfnamefont
			{H.}~\bibnamefont {Yang}}, \ and\ \bibinfo {author} {\bibfnamefont
			{J.}~\bibnamefont {Li}},\ }\href {\doibase 10.1038/nnano.2008.1} {\bibfield
		{journal} {\bibinfo  {journal} {Nature Nanotechnology}\ }\textbf {\bibinfo
			{volume} {3}},\ \bibinfo {pages} {97} (\bibinfo {year} {2008})}\BibitemShut
	{NoStop}%
	\bibitem [{\citenamefont {Parkin}, \citenamefont {Hayashi},\ and\ \citenamefont
		{Thomas}(2008)}]{Parkin08}%
	\BibitemOpen
	\bibfield  {author} {\bibinfo {author} {\bibfnamefont {S.~S.~P.}\
			\bibnamefont {Parkin}}, \bibinfo {author} {\bibfnamefont {M.}~\bibnamefont
			{Hayashi}}, \ and\ \bibinfo {author} {\bibfnamefont {L.}~\bibnamefont
			{Thomas}},\ }\href {\doibase 10.1126/science.1145799} {\bibfield  {journal}
		{\bibinfo  {journal} {Science}\ }\textbf {\bibinfo {volume} {320}},\ \bibinfo
		{pages} {190} (\bibinfo {year} {2008})}\BibitemShut {NoStop}%
	\bibitem [{\citenamefont {Woo}\ \emph {et~al.}(2016)\citenamefont {Woo},
		\citenamefont {Litzius}, \citenamefont {Krüger}, \citenamefont {Im},
		\citenamefont {Caretta}, \citenamefont {Richter}, \citenamefont {Mann},
		\citenamefont {Krone}, \citenamefont {Reeve}, \citenamefont {Weigand},
		\citenamefont {Agrawal}, \citenamefont {Lemesh}, \citenamefont {Mawass},
		\citenamefont {Fischer}, \citenamefont {Kläui},\ and\ \citenamefont
		{Beach}}]{Woo16}%
	\BibitemOpen
	\bibfield  {author} {\bibinfo {author} {\bibfnamefont {S.}~\bibnamefont
			{Woo}}, \bibinfo {author} {\bibfnamefont {K.}~\bibnamefont {Litzius}},
		\bibinfo {author} {\bibfnamefont {B.}~\bibnamefont {Krüger}}, \bibinfo
		{author} {\bibfnamefont {M.-Y.}\ \bibnamefont {Im}}, \bibinfo {author}
		{\bibfnamefont {L.}~\bibnamefont {Caretta}}, \bibinfo {author} {\bibfnamefont
			{K.}~\bibnamefont {Richter}}, \bibinfo {author} {\bibfnamefont
			{M.}~\bibnamefont {Mann}}, \bibinfo {author} {\bibfnamefont {A.}~\bibnamefont
			{Krone}}, \bibinfo {author} {\bibfnamefont {R.~M.}\ \bibnamefont {Reeve}},
		\bibinfo {author} {\bibfnamefont {M.}~\bibnamefont {Weigand}}, \bibinfo
		{author} {\bibfnamefont {P.}~\bibnamefont {Agrawal}}, \bibinfo {author}
		{\bibfnamefont {I.}~\bibnamefont {Lemesh}}, \bibinfo {author} {\bibfnamefont
			{M.-A.}\ \bibnamefont {Mawass}}, \bibinfo {author} {\bibfnamefont
			{P.}~\bibnamefont {Fischer}}, \bibinfo {author} {\bibfnamefont
			{M.}~\bibnamefont {Kläui}}, \ and\ \bibinfo {author} {\bibfnamefont
			{G.~S.~D.}\ \bibnamefont {Beach}},\ }\href {\doibase 10.1038/nmat4593}
	{\bibfield  {journal} {\bibinfo  {journal} {Nature Materials}\ }\textbf
		{\bibinfo {volume} {15}},\ \bibinfo {pages} {501} (\bibinfo {year}
		{2016})}\BibitemShut {NoStop}%
	\bibitem [{\citenamefont {Franken}, \citenamefont {Swagten},\ and\
		\citenamefont {Koopmans}(2012)}]{Franken12a}%
	\BibitemOpen
	\bibfield  {author} {\bibinfo {author} {\bibfnamefont {J.~H.}\ \bibnamefont
			{Franken}}, \bibinfo {author} {\bibfnamefont {H.~J.~M.}\ \bibnamefont
			{Swagten}}, \ and\ \bibinfo {author} {\bibfnamefont {B.}~\bibnamefont
			{Koopmans}},\ }\href {\doibase 10.1038/nnano.2012.111} {\bibfield  {journal}
		{\bibinfo  {journal} {Nature Nanotechnology}\ }\textbf {\bibinfo {volume}
			{7}},\ \bibinfo {pages} {499} (\bibinfo {year} {2012})}\BibitemShut {NoStop}%
	\bibitem [{\citenamefont {Zhang}, \citenamefont {Petford-Long},\ and\
		\citenamefont {Phatak}(2016)}]{Zhang16b}%
	\BibitemOpen
	\bibfield  {author} {\bibinfo {author} {\bibfnamefont {S.}~\bibnamefont
			{Zhang}}, \bibinfo {author} {\bibfnamefont {A.~K.}\ \bibnamefont
			{Petford-Long}}, \ and\ \bibinfo {author} {\bibfnamefont {C.}~\bibnamefont
			{Phatak}},\ }\href {\doibase 10.1038/srep31248} {\bibfield  {journal}
		{\bibinfo  {journal} {Scientific Reports}\ }\textbf {\bibinfo {volume} {6}},\
		\bibinfo {pages} {31248} (\bibinfo {year} {2016})}\BibitemShut {NoStop}%
	\bibitem [{\citenamefont {S{\'{a}}nchez-Tejerina}\ \emph
		{et~al.}(2018)\citenamefont {S{\'{a}}nchez-Tejerina}, \citenamefont
		{Mart{\'{\i}}nez}, \citenamefont {Raposo},\ and\ \citenamefont
		{Alejos}}]{Sanchez18}%
	\BibitemOpen
	\bibfield  {author} {\bibinfo {author} {\bibfnamefont {L.}~\bibnamefont
			{S{\'{a}}nchez-Tejerina}}, \bibinfo {author} {\bibfnamefont {E.}~\bibnamefont
			{Mart{\'{\i}}nez}}, \bibinfo {author} {\bibfnamefont {V.}~\bibnamefont
			{Raposo}}, \ and\ \bibinfo {author} {\bibfnamefont {{\'{O}}.}~\bibnamefont
			{Alejos}},\ }\href {\doibase 10.1063/1.4993750} {\bibfield  {journal}
		{\bibinfo  {journal} {{AIP} Advances}\ }\textbf {\bibinfo {volume} {8}},\
		\bibinfo {pages} {047302} (\bibinfo {year} {2018})}\BibitemShut {NoStop}%
	\bibitem [{\citenamefont {Ang}, \citenamefont {Gan},\ and\ \citenamefont
		{Lew}(2019)}]{Ang19}%
	\BibitemOpen
	\bibfield  {author} {\bibinfo {author} {\bibfnamefont {C.~C.~I.}\
			\bibnamefont {Ang}}, \bibinfo {author} {\bibfnamefont {W.}~\bibnamefont
			{Gan}}, \ and\ \bibinfo {author} {\bibfnamefont {W.~S.}\ \bibnamefont
			{Lew}},\ }\href {\doibase 10.1088/1367-2630/ab1171} {\bibfield  {journal}
		{\bibinfo  {journal} {New Journal of Physics}\ }\textbf {\bibinfo {volume}
			{21}},\ \bibinfo {pages} {043006} (\bibinfo {year} {2019})}\BibitemShut
	{NoStop}%
	\bibitem [{\citenamefont {D{\'{\i}}az}\ and\ \citenamefont
		{Troncoso}(2016)}]{Diaz16}%
	\BibitemOpen
	\bibfield  {author} {\bibinfo {author} {\bibfnamefont {S.~A.}\ \bibnamefont
			{D{\'{\i}}az}}\ and\ \bibinfo {author} {\bibfnamefont {R.~E.}\ \bibnamefont
			{Troncoso}},\ }\href {\doibase 10.1088/0953-8984/28/42/426005} {\bibfield
		{journal} {\bibinfo  {journal} {Journal of Physics: Condensed Matter}\
		}\textbf {\bibinfo {volume} {28}},\ \bibinfo {pages} {426005} (\bibinfo
		{year} {2016})}\BibitemShut {NoStop}%
	\bibitem [{\citenamefont {Hong}, \citenamefont {Lee},\ and\ \citenamefont
		{Lee}(2017)}]{Hong17}%
	\BibitemOpen
	\bibfield  {author} {\bibinfo {author} {\bibfnamefont {I.-S.}\ \bibnamefont
			{Hong}}, \bibinfo {author} {\bibfnamefont {S.-W.}\ \bibnamefont {Lee}}, \
		and\ \bibinfo {author} {\bibfnamefont {K.-J.}\ \bibnamefont {Lee}},\ }\href
	{\doibase 10.1016/j.cap.2017.08.024} {\bibfield  {journal} {\bibinfo
			{journal} {Current Applied Physics}\ }\textbf {\bibinfo {volume} {17}},\
		\bibinfo {pages} {1576} (\bibinfo {year} {2017})}\BibitemShut {NoStop}%
	\bibitem [{\citenamefont {Zhou}\ \emph {et~al.}(2019)\citenamefont {Zhou},
		\citenamefont {Qin}, \citenamefont {Zheng},\ and\ \citenamefont
		{Wang}}]{Zhou19}%
	\BibitemOpen
	\bibfield  {author} {\bibinfo {author} {\bibfnamefont {L.}~\bibnamefont
			{Zhou}}, \bibinfo {author} {\bibfnamefont {R.}~\bibnamefont {Qin}}, \bibinfo
		{author} {\bibfnamefont {Y.-Q.}\ \bibnamefont {Zheng}}, \ and\ \bibinfo
		{author} {\bibfnamefont {Y.}~\bibnamefont {Wang}},\ }\href {\doibase
		10.1007/s11467-019-0897-0} {\bibfield  {journal} {\bibinfo  {journal}
			{Frontiers of Physics}\ }\textbf {\bibinfo {volume} {14}} (\bibinfo {year}
		{2019}),\ 10.1007/s11467-019-0897-0}\BibitemShut {NoStop}%
	\bibitem [{\citenamefont {Menezes}\ \emph {et~al.}(2019)\citenamefont
		{Menezes}, \citenamefont {Mulkers}, \citenamefont {de~Souza~Silva},\ and\
		\citenamefont {Milo{\v{s}}evi{\'{c}}}}]{Menezes19}%
	\BibitemOpen
	\bibfield  {author} {\bibinfo {author} {\bibfnamefont {R.~M.}\ \bibnamefont
			{Menezes}}, \bibinfo {author} {\bibfnamefont {J.}~\bibnamefont {Mulkers}},
		\bibinfo {author} {\bibfnamefont {C.~C.}\ \bibnamefont {de~Souza~Silva}}, \
		and\ \bibinfo {author} {\bibfnamefont {M.~V.}\ \bibnamefont
			{Milo{\v{s}}evi{\'{c}}}},\ }\href {\doibase 10.1103/physrevb.99.104409}
	{\bibfield  {journal} {\bibinfo  {journal} {Physical Review B}\ }\textbf
		{\bibinfo {volume} {99}} (\bibinfo {year} {2019}),\
		10.1103/physrevb.99.104409}\BibitemShut {NoStop}%
	\bibitem [{\citenamefont {Toscano}\ \emph {et~al.}(2019)\citenamefont
		{Toscano}, \citenamefont {Leonel}, \citenamefont {Coura},\ and\ \citenamefont
		{Sato}}]{Toscano19}%
	\BibitemOpen
	\bibfield  {author} {\bibinfo {author} {\bibfnamefont {D.}~\bibnamefont
			{Toscano}}, \bibinfo {author} {\bibfnamefont {S.}~\bibnamefont {Leonel}},
		\bibinfo {author} {\bibfnamefont {P.}~\bibnamefont {Coura}}, \ and\ \bibinfo
		{author} {\bibfnamefont {F.}~\bibnamefont {Sato}},\ }\href {\doibase
		10.1016/j.jmmm.2019.02.075} {\bibfield  {journal} {\bibinfo  {journal}
			{Journal of Magnetism and Magnetic Materials}\ }\textbf {\bibinfo {volume}
			{480}},\ \bibinfo {pages} {171} (\bibinfo {year} {2019})}\BibitemShut
	{NoStop}%
	\bibitem [{\citenamefont {Grigoriev}\ \emph {et~al.}(2009)\citenamefont
		{Grigoriev}, \citenamefont {Chernyshov}, \citenamefont {Dyadkin},
		\citenamefont {Dmitriev}, \citenamefont {Maleyev}, \citenamefont {Moskvin},
		\citenamefont {Menzel}, \citenamefont {Schoenes},\ and\ \citenamefont
		{Eckerlebe}}]{Grigoriev09}%
	\BibitemOpen
	\bibfield  {author} {\bibinfo {author} {\bibfnamefont {S.~V.}\ \bibnamefont
			{Grigoriev}}, \bibinfo {author} {\bibfnamefont {D.}~\bibnamefont
			{Chernyshov}}, \bibinfo {author} {\bibfnamefont {V.~A.}\ \bibnamefont
			{Dyadkin}}, \bibinfo {author} {\bibfnamefont {V.}~\bibnamefont {Dmitriev}},
		\bibinfo {author} {\bibfnamefont {S.~V.}\ \bibnamefont {Maleyev}}, \bibinfo
		{author} {\bibfnamefont {E.~V.}\ \bibnamefont {Moskvin}}, \bibinfo {author}
		{\bibfnamefont {D.}~\bibnamefont {Menzel}}, \bibinfo {author} {\bibfnamefont
			{J.}~\bibnamefont {Schoenes}}, \ and\ \bibinfo {author} {\bibfnamefont
			{H.}~\bibnamefont {Eckerlebe}},\ }\href {\doibase
		10.1103/physrevlett.102.037204} {\bibfield  {journal} {\bibinfo  {journal}
			{Physical Review Letters}\ }\textbf {\bibinfo {volume} {102}} (\bibinfo
		{year} {2009}),\ 10.1103/physrevlett.102.037204}\BibitemShut {NoStop}%
	\bibitem [{\citenamefont {Grigoriev}\ \emph {et~al.}(2010)\citenamefont
		{Grigoriev}, \citenamefont {Chernyshov}, \citenamefont {Dyadkin},
		\citenamefont {Dmitriev}, \citenamefont {Moskvin}, \citenamefont {Lamago},
		\citenamefont {Wolf}, \citenamefont {Menzel}, \citenamefont {Schoenes},
		\citenamefont {Maleyev},\ and\ \citenamefont {Eckerlebe}}]{Grigoriev10}%
	\BibitemOpen
	\bibfield  {author} {\bibinfo {author} {\bibfnamefont {S.~V.}\ \bibnamefont
			{Grigoriev}}, \bibinfo {author} {\bibfnamefont {D.}~\bibnamefont
			{Chernyshov}}, \bibinfo {author} {\bibfnamefont {V.~A.}\ \bibnamefont
			{Dyadkin}}, \bibinfo {author} {\bibfnamefont {V.}~\bibnamefont {Dmitriev}},
		\bibinfo {author} {\bibfnamefont {E.~V.}\ \bibnamefont {Moskvin}}, \bibinfo
		{author} {\bibfnamefont {D.}~\bibnamefont {Lamago}}, \bibinfo {author}
		{\bibfnamefont {T.}~\bibnamefont {Wolf}}, \bibinfo {author} {\bibfnamefont
			{D.}~\bibnamefont {Menzel}}, \bibinfo {author} {\bibfnamefont
			{J.}~\bibnamefont {Schoenes}}, \bibinfo {author} {\bibfnamefont {S.~V.}\
			\bibnamefont {Maleyev}}, \ and\ \bibinfo {author} {\bibfnamefont
			{H.}~\bibnamefont {Eckerlebe}},\ }\href {\doibase 10.1103/physrevb.81.012408}
	{\bibfield  {journal} {\bibinfo  {journal} {Physical Review B}\ }\textbf
		{\bibinfo {volume} {81}} (\bibinfo {year} {2010}),\
		10.1103/physrevb.81.012408}\BibitemShut {NoStop}%
	\bibitem [{\citenamefont {Shibata}\ \emph {et~al.}(2013)\citenamefont
		{Shibata}, \citenamefont {Yu}, \citenamefont {Hara}, \citenamefont
		{Morikawa}, \citenamefont {Kanazawa}, \citenamefont {Kimoto}, \citenamefont
		{Ishiwata}, \citenamefont {Matsui},\ and\ \citenamefont
		{Tokura}}]{Shibata13}%
	\BibitemOpen
	\bibfield  {author} {\bibinfo {author} {\bibfnamefont {K.}~\bibnamefont
			{Shibata}}, \bibinfo {author} {\bibfnamefont {X.~Z.}\ \bibnamefont {Yu}},
		\bibinfo {author} {\bibfnamefont {T.}~\bibnamefont {Hara}}, \bibinfo {author}
		{\bibfnamefont {D.}~\bibnamefont {Morikawa}}, \bibinfo {author}
		{\bibfnamefont {N.}~\bibnamefont {Kanazawa}}, \bibinfo {author}
		{\bibfnamefont {K.}~\bibnamefont {Kimoto}}, \bibinfo {author} {\bibfnamefont
			{S.}~\bibnamefont {Ishiwata}}, \bibinfo {author} {\bibfnamefont
			{Y.}~\bibnamefont {Matsui}}, \ and\ \bibinfo {author} {\bibfnamefont
			{Y.}~\bibnamefont {Tokura}},\ }\href {\doibase 10.1038/nnano.2013.174}
	{\bibfield  {journal} {\bibinfo  {journal} {Nature Nanotechnology}\ }\textbf
		{\bibinfo {volume} {8}},\ \bibinfo {pages} {723} (\bibinfo {year}
		{2013})}\BibitemShut {NoStop}%
	\bibitem [{\citenamefont {Grigoriev}\ \emph {et~al.}(2013)\citenamefont
		{Grigoriev}, \citenamefont {Potapova}, \citenamefont {Siegfried},
		\citenamefont {Dyadkin}, \citenamefont {Moskvin}, \citenamefont {Dmitriev},
		\citenamefont {Menzel}, \citenamefont {Dewhurst}, \citenamefont {Chernyshov},
		\citenamefont {Sadykov}, \citenamefont {Fomicheva},\ and\ \citenamefont
		{Tsvyashchenko}}]{Grigoriev13}%
	\BibitemOpen
	\bibfield  {author} {\bibinfo {author} {\bibfnamefont {S.~V.}\ \bibnamefont
			{Grigoriev}}, \bibinfo {author} {\bibfnamefont {N.~M.}\ \bibnamefont
			{Potapova}}, \bibinfo {author} {\bibfnamefont {S.-A.}\ \bibnamefont
			{Siegfried}}, \bibinfo {author} {\bibfnamefont {V.~A.}\ \bibnamefont
			{Dyadkin}}, \bibinfo {author} {\bibfnamefont {E.~V.}\ \bibnamefont
			{Moskvin}}, \bibinfo {author} {\bibfnamefont {V.}~\bibnamefont {Dmitriev}},
		\bibinfo {author} {\bibfnamefont {D.}~\bibnamefont {Menzel}}, \bibinfo
		{author} {\bibfnamefont {C.~D.}\ \bibnamefont {Dewhurst}}, \bibinfo {author}
		{\bibfnamefont {D.}~\bibnamefont {Chernyshov}}, \bibinfo {author}
		{\bibfnamefont {R.~A.}\ \bibnamefont {Sadykov}}, \bibinfo {author}
		{\bibfnamefont {L.~N.}\ \bibnamefont {Fomicheva}}, \ and\ \bibinfo {author}
		{\bibfnamefont {A.~V.}\ \bibnamefont {Tsvyashchenko}},\ }\href {\doibase
		10.1103/physrevlett.110.207201} {\bibfield  {journal} {\bibinfo  {journal}
			{Physical Review Letters}\ }\textbf {\bibinfo {volume} {110}} (\bibinfo
		{year} {2013}),\ 10.1103/physrevlett.110.207201}\BibitemShut {NoStop}%
	\bibitem [{\citenamefont {Morikawa}\ \emph {et~al.}(2013)\citenamefont
		{Morikawa}, \citenamefont {Shibata}, \citenamefont {Kanazawa}, \citenamefont
		{Yu},\ and\ \citenamefont {Tokura}}]{Morikawa13}%
	\BibitemOpen
	\bibfield  {author} {\bibinfo {author} {\bibfnamefont {D.}~\bibnamefont
			{Morikawa}}, \bibinfo {author} {\bibfnamefont {K.}~\bibnamefont {Shibata}},
		\bibinfo {author} {\bibfnamefont {N.}~\bibnamefont {Kanazawa}}, \bibinfo
		{author} {\bibfnamefont {X.~Z.}\ \bibnamefont {Yu}}, \ and\ \bibinfo {author}
		{\bibfnamefont {Y.}~\bibnamefont {Tokura}},\ }\href {\doibase
		10.1103/physrevb.88.024408} {\bibfield  {journal} {\bibinfo  {journal}
			{Physical Review B}\ }\textbf {\bibinfo {volume} {88}} (\bibinfo {year}
		{2013}),\ 10.1103/physrevb.88.024408}\BibitemShut {NoStop}%
	\bibitem [{\citenamefont {Siegfried}\ \emph {et~al.}(2015)\citenamefont
		{Siegfried}, \citenamefont {Altynbaev}, \citenamefont {Chubova},
		\citenamefont {Dyadkin}, \citenamefont {Chernyshov}, \citenamefont {Moskvin},
		\citenamefont {Menzel}, \citenamefont {Heinemann}, \citenamefont {Schreyer},\
		and\ \citenamefont {Grigoriev}}]{Siegfried15}%
	\BibitemOpen
	\bibfield  {author} {\bibinfo {author} {\bibfnamefont {S.-A.}\ \bibnamefont
			{Siegfried}}, \bibinfo {author} {\bibfnamefont {E.~V.}\ \bibnamefont
			{Altynbaev}}, \bibinfo {author} {\bibfnamefont {N.~M.}\ \bibnamefont
			{Chubova}}, \bibinfo {author} {\bibfnamefont {V.}~\bibnamefont {Dyadkin}},
		\bibinfo {author} {\bibfnamefont {D.}~\bibnamefont {Chernyshov}}, \bibinfo
		{author} {\bibfnamefont {E.~V.}\ \bibnamefont {Moskvin}}, \bibinfo {author}
		{\bibfnamefont {D.}~\bibnamefont {Menzel}}, \bibinfo {author} {\bibfnamefont
			{A.}~\bibnamefont {Heinemann}}, \bibinfo {author} {\bibfnamefont
			{A.}~\bibnamefont {Schreyer}}, \ and\ \bibinfo {author} {\bibfnamefont
			{S.~V.}\ \bibnamefont {Grigoriev}},\ }\href {\doibase
		10.1103/physrevb.91.184406} {\bibfield  {journal} {\bibinfo  {journal}
			{Physical Review B}\ }\textbf {\bibinfo {volume} {91}} (\bibinfo {year}
		{2015}),\ 10.1103/physrevb.91.184406}\BibitemShut {NoStop}%
	\bibitem [{\citenamefont {Koretsune}, \citenamefont {Nagaosa},\ and\
		\citenamefont {Arita}(2015)}]{Koretsune15}%
	\BibitemOpen
	\bibfield  {author} {\bibinfo {author} {\bibfnamefont {T.}~\bibnamefont
			{Koretsune}}, \bibinfo {author} {\bibfnamefont {N.}~\bibnamefont {Nagaosa}},
		\ and\ \bibinfo {author} {\bibfnamefont {R.}~\bibnamefont {Arita}},\ }\href
	{\doibase 10.1038/srep13302} {\bibfield  {journal} {\bibinfo  {journal}
			{Scientific Reports}\ }\textbf {\bibinfo {volume} {5}} (\bibinfo {year}
		{2015}),\ 10.1038/srep13302}\BibitemShut {NoStop}%
	\bibitem [{\citenamefont {Ma}\ \emph {et~al.}(2016)\citenamefont {Ma},
		\citenamefont {Yu}, \citenamefont {Li}, \citenamefont {Wang}, \citenamefont
		{Wu}, \citenamefont {Olsson}, \citenamefont {Chu}, \citenamefont {An},
		\citenamefont {Xiao}, \citenamefont {Wang},\ and\ \citenamefont
		{Li}}]{Ma16a}%
	\BibitemOpen
	\bibfield  {author} {\bibinfo {author} {\bibfnamefont {X.}~\bibnamefont
			{Ma}}, \bibinfo {author} {\bibfnamefont {G.}~\bibnamefont {Yu}}, \bibinfo
		{author} {\bibfnamefont {X.}~\bibnamefont {Li}}, \bibinfo {author}
		{\bibfnamefont {T.}~\bibnamefont {Wang}}, \bibinfo {author} {\bibfnamefont
			{D.}~\bibnamefont {Wu}}, \bibinfo {author} {\bibfnamefont {K.~S.}\
			\bibnamefont {Olsson}}, \bibinfo {author} {\bibfnamefont {Z.}~\bibnamefont
			{Chu}}, \bibinfo {author} {\bibfnamefont {K.}~\bibnamefont {An}}, \bibinfo
		{author} {\bibfnamefont {J.~Q.}\ \bibnamefont {Xiao}}, \bibinfo {author}
		{\bibfnamefont {K.~L.}\ \bibnamefont {Wang}}, \ and\ \bibinfo {author}
		{\bibfnamefont {X.}~\bibnamefont {Li}},\ }\href {\doibase
		10.1103/physrevb.94.180408} {\bibfield  {journal} {\bibinfo  {journal}
			{Physical Review B}\ }\textbf {\bibinfo {volume} {94}} (\bibinfo {year}
		{2016}),\ 10.1103/physrevb.94.180408}\BibitemShut {NoStop}%
	\bibitem [{\citenamefont {Tacchi}\ \emph {et~al.}(2017)\citenamefont {Tacchi},
		\citenamefont {Troncoso}, \citenamefont {Ahlberg}, \citenamefont {Gubbiotti},
		\citenamefont {Madami}, \citenamefont {Akerman},\ and\ \citenamefont
		{Landeros}}]{Tacchi17}%
	\BibitemOpen
	\bibfield  {author} {\bibinfo {author} {\bibfnamefont {S.}~\bibnamefont
			{Tacchi}}, \bibinfo {author} {\bibfnamefont {R.~E.}\ \bibnamefont
			{Troncoso}}, \bibinfo {author} {\bibfnamefont {M.}~\bibnamefont {Ahlberg}},
		\bibinfo {author} {\bibfnamefont {G.}~\bibnamefont {Gubbiotti}}, \bibinfo
		{author} {\bibfnamefont {M.}~\bibnamefont {Madami}}, \bibinfo {author}
		{\bibfnamefont {J.}~\bibnamefont {Akerman}}, \ and\ \bibinfo {author}
		{\bibfnamefont {P.}~\bibnamefont {Landeros}},\ }\href
	{http://www.arXiv.org/abs/1604.02626} {\bibfield  {journal} {\bibinfo
			{journal} {Physical Review Letters}\ } (\bibinfo {year} {2017})}\BibitemShut
	{NoStop}%
	\bibitem [{\citenamefont {Nembach}\ \emph {et~al.}(2015)\citenamefont
		{Nembach}, \citenamefont {Shaw}, \citenamefont {Weiler}, \citenamefont
		{Ju{\'{e}}},\ and\ \citenamefont {Silva}}]{Nembach15}%
	\BibitemOpen
	\bibfield  {author} {\bibinfo {author} {\bibfnamefont {H.~T.}\ \bibnamefont
			{Nembach}}, \bibinfo {author} {\bibfnamefont {J.~M.}\ \bibnamefont {Shaw}},
		\bibinfo {author} {\bibfnamefont {M.}~\bibnamefont {Weiler}}, \bibinfo
		{author} {\bibfnamefont {E.}~\bibnamefont {Ju{\'{e}}}}, \ and\ \bibinfo
		{author} {\bibfnamefont {T.~J.}\ \bibnamefont {Silva}},\ }\href {\doibase
		10.1038/nphys3418} {\bibfield  {journal} {\bibinfo  {journal} {Nature
				Physics}\ }\textbf {\bibinfo {volume} {11}},\ \bibinfo {pages} {825}
		(\bibinfo {year} {2015})}\BibitemShut {NoStop}%
	\bibitem [{\citenamefont {Belmeguenai}\ \emph {et~al.}(2015)\citenamefont
		{Belmeguenai}, \citenamefont {Adam}, \citenamefont {Roussign\'e},
		\citenamefont {Eimer}, \citenamefont {Devolder}, \citenamefont {Kim},
		\citenamefont {Cherif}, \citenamefont {Stashkevich},\ and\ \citenamefont
		{Thiaville}}]{Belmeguenai15}%
	\BibitemOpen
	\bibfield  {author} {\bibinfo {author} {\bibfnamefont {M.}~\bibnamefont
			{Belmeguenai}}, \bibinfo {author} {\bibfnamefont {J.-P.}\ \bibnamefont
			{Adam}}, \bibinfo {author} {\bibfnamefont {Y.}~\bibnamefont {Roussign\'e}},
		\bibinfo {author} {\bibfnamefont {S.}~\bibnamefont {Eimer}}, \bibinfo
		{author} {\bibfnamefont {T.}~\bibnamefont {Devolder}}, \bibinfo {author}
		{\bibfnamefont {J.-V.}\ \bibnamefont {Kim}}, \bibinfo {author} {\bibfnamefont
			{S.~M.}\ \bibnamefont {Cherif}}, \bibinfo {author} {\bibfnamefont
			{A.}~\bibnamefont {Stashkevich}}, \ and\ \bibinfo {author} {\bibfnamefont
			{A.}~\bibnamefont {Thiaville}},\ }\href {\doibase 10.1103/PhysRevB.91.180405}
	{\bibfield  {journal} {\bibinfo  {journal} {Physical Review B}\ }\textbf
		{\bibinfo {volume} {91}},\ \bibinfo {pages} {180405} (\bibinfo {year}
		{2015})}\BibitemShut {NoStop}%
	\bibitem [{\citenamefont {Stashkevich}\ \emph {et~al.}(2015)\citenamefont
		{Stashkevich}, \citenamefont {Belmeguenai}, \citenamefont {Roussign{\'{e}}},
		\citenamefont {Cherif}, \citenamefont {Kostylev}, \citenamefont {Gabor},
		\citenamefont {Lacour}, \citenamefont {Tiusan},\ and\ \citenamefont
		{Hehn}}]{Stashkevich15}%
	\BibitemOpen
	\bibfield  {author} {\bibinfo {author} {\bibfnamefont {A.~A.}\ \bibnamefont
			{Stashkevich}}, \bibinfo {author} {\bibfnamefont {M.}~\bibnamefont
			{Belmeguenai}}, \bibinfo {author} {\bibfnamefont {Y.}~\bibnamefont
			{Roussign{\'{e}}}}, \bibinfo {author} {\bibfnamefont {S.~M.}\ \bibnamefont
			{Cherif}}, \bibinfo {author} {\bibfnamefont {M.}~\bibnamefont {Kostylev}},
		\bibinfo {author} {\bibfnamefont {M.}~\bibnamefont {Gabor}}, \bibinfo
		{author} {\bibfnamefont {D.}~\bibnamefont {Lacour}}, \bibinfo {author}
		{\bibfnamefont {C.}~\bibnamefont {Tiusan}}, \ and\ \bibinfo {author}
		{\bibfnamefont {M.}~\bibnamefont {Hehn}},\ }\href {\doibase
		10.1103/physrevb.91.214409} {\bibfield  {journal} {\bibinfo  {journal}
			{Physical Review B}\ }\textbf {\bibinfo {volume} {91}} (\bibinfo {year}
		{2015}),\ 10.1103/physrevb.91.214409}\BibitemShut {NoStop}%
	\bibitem [{\citenamefont {Lee}\ \emph {et~al.}(2015)\citenamefont {Lee},
		\citenamefont {Jang}, \citenamefont {Min}, \citenamefont {Lee}, \citenamefont
		{Lee},\ and\ \citenamefont {Chang}}]{Lee15a}%
	\BibitemOpen
	\bibfield  {author} {\bibinfo {author} {\bibfnamefont {J.~M.}\ \bibnamefont
			{Lee}}, \bibinfo {author} {\bibfnamefont {C.}~\bibnamefont {Jang}}, \bibinfo
		{author} {\bibfnamefont {B.-C.}\ \bibnamefont {Min}}, \bibinfo {author}
		{\bibfnamefont {S.-W.}\ \bibnamefont {Lee}}, \bibinfo {author} {\bibfnamefont
			{K.-J.}\ \bibnamefont {Lee}}, \ and\ \bibinfo {author} {\bibfnamefont
			{J.}~\bibnamefont {Chang}},\ }\href {\doibase 10.1021/acs.nanolett.5b02732}
	{\bibfield  {journal} {\bibinfo  {journal} {Nano Letters}\ }\textbf {\bibinfo
			{volume} {16}},\ \bibinfo {pages} {62} (\bibinfo {year} {2015})}\BibitemShut
	{NoStop}%
	\bibitem [{\citenamefont {Kopte}\ \emph {et~al.}(2017)\citenamefont {Kopte},
		\citenamefont {R{\"o}{\ss}ler}, \citenamefont {Sch{\"a}fer}, \citenamefont
		{Kosub}, \citenamefont {K{\'a}kay}, \citenamefont {Volkov}, \citenamefont
		{Fuchs}, \citenamefont {Vedmedenko}, \citenamefont {Radu}, \citenamefont
		{Schmidt}, \citenamefont {Lindner}, \citenamefont {Fa{\ss}bender},\ and\
		\citenamefont {Makarov}}]{Kopte17}%
	\BibitemOpen
	\bibfield  {author} {\bibinfo {author} {\bibfnamefont {M.}~\bibnamefont
			{Kopte}}, \bibinfo {author} {\bibfnamefont {U.~K.}\ \bibnamefont
			{R{\"o}{\ss}ler}}, \bibinfo {author} {\bibfnamefont {R.}~\bibnamefont
			{Sch{\"a}fer}}, \bibinfo {author} {\bibfnamefont {T.}~\bibnamefont {Kosub}},
		\bibinfo {author} {\bibfnamefont {A.}~\bibnamefont {K{\'a}kay}}, \bibinfo
		{author} {\bibfnamefont {O.}~\bibnamefont {Volkov}}, \bibinfo {author}
		{\bibfnamefont {H.}~\bibnamefont {Fuchs}}, \bibinfo {author} {\bibfnamefont
			{E.~Y.}\ \bibnamefont {Vedmedenko}}, \bibinfo {author} {\bibfnamefont
			{F.}~\bibnamefont {Radu}}, \bibinfo {author} {\bibfnamefont {O.~G.}\
			\bibnamefont {Schmidt}}, \bibinfo {author} {\bibfnamefont {J.}~\bibnamefont
			{Lindner}}, \bibinfo {author} {\bibfnamefont {J.}~\bibnamefont
			{Fa{\ss}bender}}, \ and\ \bibinfo {author} {\bibfnamefont {D.}~\bibnamefont
			{Makarov}},\ }\href {https://arxiv.org/abs/1706.09322} {\bibfield  {journal}
		{\bibinfo  {journal} {ArXiv e-prints}\ } (\bibinfo {year} {2017})},\ \Eprint
	{http://arxiv.org/abs/1706.09322v1} {1706.09322v1} \BibitemShut {NoStop}%
	\bibitem [{\citenamefont {Srivastava}\ \emph {et~al.}(2018)\citenamefont
		{Srivastava}, \citenamefont {Schott}, \citenamefont {Juge}, \citenamefont
		{K{\v{r}}i{\v{z}}{\'{a}}kov{\'{a}}}, \citenamefont {Belmeguenai},
		\citenamefont {Roussign{\'{e}}}, \citenamefont {Bernand-Mantel},
		\citenamefont {Ranno}, \citenamefont {Pizzini}, \citenamefont {Ch{\'{e}}rif},
		\citenamefont {Stashkevich}, \citenamefont {Auffret}, \citenamefont {Boulle},
		\citenamefont {Gaudin}, \citenamefont {Chshiev}, \citenamefont {Baraduc},\
		and\ \citenamefont {B{\'{e}}a}}]{Srivastava18}%
	\BibitemOpen
	\bibfield  {author} {\bibinfo {author} {\bibfnamefont {T.}~\bibnamefont
			{Srivastava}}, \bibinfo {author} {\bibfnamefont {M.}~\bibnamefont {Schott}},
		\bibinfo {author} {\bibfnamefont {R.}~\bibnamefont {Juge}}, \bibinfo {author}
		{\bibfnamefont {V.}~\bibnamefont {K{\v{r}}i{\v{z}}{\'{a}}kov{\'{a}}}},
		\bibinfo {author} {\bibfnamefont {M.}~\bibnamefont {Belmeguenai}}, \bibinfo
		{author} {\bibfnamefont {Y.}~\bibnamefont {Roussign{\'{e}}}}, \bibinfo
		{author} {\bibfnamefont {A.}~\bibnamefont {Bernand-Mantel}}, \bibinfo
		{author} {\bibfnamefont {L.}~\bibnamefont {Ranno}}, \bibinfo {author}
		{\bibfnamefont {S.}~\bibnamefont {Pizzini}}, \bibinfo {author} {\bibfnamefont
			{S.-M.}\ \bibnamefont {Ch{\'{e}}rif}}, \bibinfo {author} {\bibfnamefont
			{A.}~\bibnamefont {Stashkevich}}, \bibinfo {author} {\bibfnamefont
			{S.}~\bibnamefont {Auffret}}, \bibinfo {author} {\bibfnamefont
			{O.}~\bibnamefont {Boulle}}, \bibinfo {author} {\bibfnamefont
			{G.}~\bibnamefont {Gaudin}}, \bibinfo {author} {\bibfnamefont
			{M.}~\bibnamefont {Chshiev}}, \bibinfo {author} {\bibfnamefont
			{C.}~\bibnamefont {Baraduc}}, \ and\ \bibinfo {author} {\bibfnamefont
			{H.}~\bibnamefont {B{\'{e}}a}},\ }\href {\doibase
		10.1021/acs.nanolett.8b01502} {\bibfield  {journal} {\bibinfo  {journal}
			{Nano Letters}\ }\textbf {\bibinfo {volume} {18}},\ \bibinfo {pages} {4871}
		(\bibinfo {year} {2018})}\BibitemShut {NoStop}%
	\bibitem [{\citenamefont {Yang}\ \emph {et~al.}(2018)\citenamefont {Yang},
		\citenamefont {Boulle}, \citenamefont {Cros}, \citenamefont {Fert},\ and\
		\citenamefont {Chshiev}}]{Yang18b}%
	\BibitemOpen
	\bibfield  {author} {\bibinfo {author} {\bibfnamefont {H.}~\bibnamefont
			{Yang}}, \bibinfo {author} {\bibfnamefont {O.}~\bibnamefont {Boulle}},
		\bibinfo {author} {\bibfnamefont {V.}~\bibnamefont {Cros}}, \bibinfo {author}
		{\bibfnamefont {A.}~\bibnamefont {Fert}}, \ and\ \bibinfo {author}
		{\bibfnamefont {M.}~\bibnamefont {Chshiev}},\ }\href {\doibase
		10.1038/s41598-018-30063-y} {\bibfield  {journal} {\bibinfo  {journal}
			{Scientific Reports}\ }\textbf {\bibinfo {volume} {8}},\ \bibinfo {pages}
		{12356} (\bibinfo {year} {2018})}\BibitemShut {NoStop}%
	\bibitem [{\citenamefont {Balk}\ \emph {et~al.}(2017)\citenamefont {Balk},
		\citenamefont {Kim}, \citenamefont {Pierce}, \citenamefont {Stiles},
		\citenamefont {Unguris},\ and\ \citenamefont {Stavis}}]{Balk17}%
	\BibitemOpen
	\bibfield  {author} {\bibinfo {author} {\bibfnamefont {A.}~\bibnamefont
			{Balk}}, \bibinfo {author} {\bibfnamefont {K.-W.}\ \bibnamefont {Kim}},
		\bibinfo {author} {\bibfnamefont {D.}~\bibnamefont {Pierce}}, \bibinfo
		{author} {\bibfnamefont {M.}~\bibnamefont {Stiles}}, \bibinfo {author}
		{\bibfnamefont {J.}~\bibnamefont {Unguris}}, \ and\ \bibinfo {author}
		{\bibfnamefont {S.}~\bibnamefont {Stavis}},\ }\href {\doibase
		10.1103/physrevlett.119.077205} {\bibfield  {journal} {\bibinfo  {journal}
			{Physical Review Letters}\ }\textbf {\bibinfo {volume} {119}} (\bibinfo
		{year} {2017}),\ 10.1103/physrevlett.119.077205}\BibitemShut {NoStop}%
	\bibitem [{\citenamefont {Yershov}\ \emph {et~al.}(2015)\citenamefont
		{Yershov}, \citenamefont {Kravchuk}, \citenamefont {Sheka},\ and\
		\citenamefont {Gaididei}}]{Yershov15b}%
	\BibitemOpen
	\bibfield  {author} {\bibinfo {author} {\bibfnamefont {K.~V.}\ \bibnamefont
			{Yershov}}, \bibinfo {author} {\bibfnamefont {V.~P.}\ \bibnamefont
			{Kravchuk}}, \bibinfo {author} {\bibfnamefont {D.~D.}\ \bibnamefont {Sheka}},
		\ and\ \bibinfo {author} {\bibfnamefont {Y.}~\bibnamefont {Gaididei}},\
	}\href {\doibase 10.1103/PhysRevB.92.104412} {\bibfield  {journal} {\bibinfo
			{journal} {Physical Review B}\ }\textbf {\bibinfo {volume} {92}},\ \bibinfo
		{pages} {104412} (\bibinfo {year} {2015})}\BibitemShut {NoStop}%
	\bibitem [{\citenamefont {Yershov}\ \emph {et~al.}(2018)\citenamefont
		{Yershov}, \citenamefont {Kravchuk}, \citenamefont {Sheka}, \citenamefont
		{Pylypovskyi}, \citenamefont {Makarov},\ and\ \citenamefont
		{Gaididei}}]{Yershov18a}%
	\BibitemOpen
	\bibfield  {author} {\bibinfo {author} {\bibfnamefont {K.~V.}\ \bibnamefont
			{Yershov}}, \bibinfo {author} {\bibfnamefont {V.~P.}\ \bibnamefont
			{Kravchuk}}, \bibinfo {author} {\bibfnamefont {D.~D.}\ \bibnamefont {Sheka}},
		\bibinfo {author} {\bibfnamefont {O.~V.}\ \bibnamefont {Pylypovskyi}},
		\bibinfo {author} {\bibfnamefont {D.}~\bibnamefont {Makarov}}, \ and\
		\bibinfo {author} {\bibfnamefont {Y.}~\bibnamefont {Gaididei}},\ }\href
	{\doibase 10.1103/physrevb.98.060409} {\bibfield  {journal} {\bibinfo
			{journal} {Physical Review B}\ }\textbf {\bibinfo {volume} {98}},\ \bibinfo
		{pages} {060409} (\bibinfo {year} {2018})}\BibitemShut {NoStop}%
	\bibitem [{\citenamefont {Bazaliy}, \citenamefont {Jones},\ and\ \citenamefont
		{Zhang}(1998)}]{Bazaliy98}%
	\BibitemOpen
	\bibfield  {author} {\bibinfo {author} {\bibfnamefont {Y.~B.}\ \bibnamefont
			{Bazaliy}}, \bibinfo {author} {\bibfnamefont {B.~A.}\ \bibnamefont {Jones}},
		\ and\ \bibinfo {author} {\bibfnamefont {S.-C.}\ \bibnamefont {Zhang}},\
	}\href {\doibase 10.1103/PhysRevB.57.R3213} {\bibfield  {journal} {\bibinfo
			{journal} {Physical Review B}\ }\textbf {\bibinfo {volume} {57}},\ \bibinfo
		{pages} {R3213} (\bibinfo {year} {1998})}\BibitemShut {NoStop}%
	\bibitem [{\citenamefont {Zhang}\ and\ \citenamefont {Li}(2004)}]{Zhang04}%
	\BibitemOpen
	\bibfield  {author} {\bibinfo {author} {\bibfnamefont {S.}~\bibnamefont
			{Zhang}}\ and\ \bibinfo {author} {\bibfnamefont {Z.}~\bibnamefont {Li}},\
	}\href {\doibase 10.1103/PhysRevLett.93.127204} {\bibfield  {journal}
		{\bibinfo  {journal} {Physical Review Letters}\ }\textbf {\bibinfo {volume}
			{93}},\ \bibinfo {pages} {127204} (\bibinfo {year} {2004})}\BibitemShut
	{NoStop}%
	\bibitem [{\citenamefont {Thiaville}\ \emph {et~al.}(2005)\citenamefont
		{Thiaville}, \citenamefont {Nakatani}, \citenamefont {Miltat},\ and\
		\citenamefont {Suzuki}}]{Thiaville05}%
	\BibitemOpen
	\bibfield  {author} {\bibinfo {author} {\bibfnamefont {A.}~\bibnamefont
			{Thiaville}}, \bibinfo {author} {\bibfnamefont {Y.}~\bibnamefont {Nakatani}},
		\bibinfo {author} {\bibfnamefont {J.}~\bibnamefont {Miltat}}, \ and\ \bibinfo
		{author} {\bibfnamefont {Y.}~\bibnamefont {Suzuki}},\ }\href
	{http://stacks.iop.org/0295-5075/69/990} {\bibfield  {journal} {\bibinfo
			{journal} {Europhysics Letters (EPL)}\ }\textbf {\bibinfo {volume} {69}},\
		\bibinfo {pages} {990} (\bibinfo {year} {2005})}\BibitemShut {NoStop}%
	\bibitem [{\citenamefont {Aharoni}(1998)}]{Aharoni98}%
	\BibitemOpen
	\bibfield  {author} {\bibinfo {author} {\bibfnamefont {A.}~\bibnamefont
			{Aharoni}},\ }\href {\doibase 10.1063/1.367113} {\bibfield  {journal}
		{\bibinfo  {journal} {Journal of Applied Physics}\ }\textbf {\bibinfo
			{volume} {83}},\ \bibinfo {pages} {3432} (\bibinfo {year}
		{1998})}\BibitemShut {NoStop}%
	\bibitem [{\citenamefont {Porter}\ and\ \citenamefont
		{Donahue}(2004)}]{Porter04}%
	\BibitemOpen
	\bibfield  {author} {\bibinfo {author} {\bibfnamefont {D.~G.}\ \bibnamefont
			{Porter}}\ and\ \bibinfo {author} {\bibfnamefont {M.~J.}\ \bibnamefont
			{Donahue}},\ }\href {\doibase 10.1063/1.1688673} {\bibfield  {journal}
		{\bibinfo  {journal} {Journal of Applied Physics}\ }\textbf {\bibinfo
			{volume} {95}},\ \bibinfo {pages} {6729} (\bibinfo {year}
		{2004})}\BibitemShut {NoStop}%
	\bibitem [{\citenamefont {Hillebrands}\ and\ \citenamefont
		{Thiaville}(2006)}]{Hillebrands06}%
	\BibitemOpen
	\bibinfo {editor} {\bibfnamefont {B.}~\bibnamefont {Hillebrands}}\ and\
	\bibinfo {editor} {\bibfnamefont {A.}~\bibnamefont {Thiaville}},\ eds.,\
	\href@noop {} {\emph {\bibinfo {title} {Spin dynamics in confined magnetic
				structures III}}},\ \bibinfo {series} {Topics in Applied Physics}, Vol.\
	\bibinfo {volume} {101}\ (\bibinfo  {publisher} {Springer},\ \bibinfo
	{address} {Berlin},\ \bibinfo {year} {2006})\BibitemShut {NoStop}%
	\bibitem [{\citenamefont {Mougin}\ \emph {et~al.}(2007)\citenamefont {Mougin},
		\citenamefont {Cormier}, \citenamefont {Adam}, \citenamefont {Metaxas},\ and\
		\citenamefont {Ferr{\'e}}}]{Mougin07}%
	\BibitemOpen
	\bibfield  {author} {\bibinfo {author} {\bibfnamefont {A.}~\bibnamefont
			{Mougin}}, \bibinfo {author} {\bibfnamefont {M.}~\bibnamefont {Cormier}},
		\bibinfo {author} {\bibfnamefont {J.~P.}\ \bibnamefont {Adam}}, \bibinfo
		{author} {\bibfnamefont {P.~J.}\ \bibnamefont {Metaxas}}, \ and\ \bibinfo
		{author} {\bibfnamefont {J.}~\bibnamefont {Ferr{\'e}}},\ }\href
	{http://stacks.iop.org/0295-5075/78/i=5/a=57007} {\bibfield  {journal}
		{\bibinfo  {journal} {EPL (Europhysics Letters)}\ }\textbf {\bibinfo {volume}
			{78}},\ \bibinfo {pages} {57007} (\bibinfo {year} {2007})}\BibitemShut
	{NoStop}%
	\bibitem [{\citenamefont {Bogdanov}\ and\ \citenamefont
		{R{\"o}{\ss}ler}(2001)}]{Bogdanov01}%
	\BibitemOpen
	\bibfield  {author} {\bibinfo {author} {\bibfnamefont {A.}~\bibnamefont
			{Bogdanov}}\ and\ \bibinfo {author} {\bibfnamefont {U.}~\bibnamefont
			{R{\"o}{\ss}ler}},\ }\href {\doibase 10.1103/physrevlett.87.037203}
	{\bibfield  {journal} {\bibinfo  {journal} {Physical Review Letters}\
		}\textbf {\bibinfo {volume} {87}},\ \bibinfo {pages} {037203} (\bibinfo
		{year} {2001})}\BibitemShut {NoStop}%
	\bibitem [{\citenamefont {Thiaville}\ \emph {et~al.}(2012)\citenamefont
		{Thiaville}, \citenamefont {Rohart}, \citenamefont {Ju{\'e}}, \citenamefont
		{Cros},\ and\ \citenamefont {Fert}}]{Thiaville12}%
	\BibitemOpen
	\bibfield  {author} {\bibinfo {author} {\bibfnamefont {A.}~\bibnamefont
			{Thiaville}}, \bibinfo {author} {\bibfnamefont {S.}~\bibnamefont {Rohart}},
		\bibinfo {author} {\bibfnamefont {{\'E}.}~\bibnamefont {Ju{\'e}}}, \bibinfo
		{author} {\bibfnamefont {V.}~\bibnamefont {Cros}}, \ and\ \bibinfo {author}
		{\bibfnamefont {A.}~\bibnamefont {Fert}},\ }\href {\doibase
		10.1209/0295-5075/100/57002} {\bibfield  {journal} {\bibinfo  {journal} {EPL
				(Europhysics Letters)}\ }\textbf {\bibinfo {volume} {100}},\ \bibinfo {pages}
		{57002} (\bibinfo {year} {2012})}\BibitemShut {NoStop}%
	\bibitem [{\citenamefont {Yang}\ \emph {et~al.}(2015)\citenamefont {Yang},
		\citenamefont {Thiaville}, \citenamefont {Rohart}, \citenamefont {Fert},\
		and\ \citenamefont {Chshiev}}]{Yang15}%
	\BibitemOpen
	\bibfield  {author} {\bibinfo {author} {\bibfnamefont {H.}~\bibnamefont
			{Yang}}, \bibinfo {author} {\bibfnamefont {A.}~\bibnamefont {Thiaville}},
		\bibinfo {author} {\bibfnamefont {S.}~\bibnamefont {Rohart}}, \bibinfo
		{author} {\bibfnamefont {A.}~\bibnamefont {Fert}}, \ and\ \bibinfo {author}
		{\bibfnamefont {M.}~\bibnamefont {Chshiev}},\ }\href {\doibase
		10.1103/PhysRevLett.115.267210} {\bibfield  {journal} {\bibinfo  {journal}
			{Physical Review Letters}\ }\textbf {\bibinfo {volume} {115}},\ \bibinfo
		{pages} {267210} (\bibinfo {year} {2015})}\BibitemShut {NoStop}%
	\bibitem [{\citenamefont {Leonov}\ \emph {et~al.}(2016)\citenamefont {Leonov},
		\citenamefont {Monchesky}, \citenamefont {Romming}, \citenamefont {Kubetzka},
		\citenamefont {Bogdanov},\ and\ \citenamefont {Wiesendanger}}]{Leonov16}%
	\BibitemOpen
	\bibfield  {author} {\bibinfo {author} {\bibfnamefont {A.~O.}\ \bibnamefont
			{Leonov}}, \bibinfo {author} {\bibfnamefont {T.~L.}\ \bibnamefont
			{Monchesky}}, \bibinfo {author} {\bibfnamefont {N.}~\bibnamefont {Romming}},
		\bibinfo {author} {\bibfnamefont {A.}~\bibnamefont {Kubetzka}}, \bibinfo
		{author} {\bibfnamefont {A.~N.}\ \bibnamefont {Bogdanov}}, \ and\ \bibinfo
		{author} {\bibfnamefont {R.}~\bibnamefont {Wiesendanger}},\ }\href {\doibase
		10.1088/1367-2630/18/6/065003} {\bibfield  {journal} {\bibinfo  {journal}
			{New Journal of Physics}\ }\textbf {\bibinfo {volume} {18}},\ \bibinfo
		{pages} {065003} (\bibinfo {year} {2016})}\BibitemShut {NoStop}%
	\bibitem [{Note1()}]{Note1}%
	\BibitemOpen
	\bibinfo {note} {Here we consider the spatial distribution of DMI strength
		without the change of sign, i.e. $d\geq 0$.}\BibitemShut {Stop}%
	\bibitem [{Note2()}]{Note2}%
	\BibitemOpen
	\bibinfo {note} {See Supplemental Material at \protect \texttt {Link provided
			by the publisher} for details of analytical calculations and movies, which
		includes Refs.~\cite
		{Bazaliy98,Zhang04,Thiaville05,Doering48,Hillebrands06,Mougin07,Yershov16}.}\BibitemShut
	{Stop}%
	\bibitem [{\citenamefont {Slonczewski}(1972)}]{Slonczewski72}%
	\BibitemOpen
	\bibfield  {author} {\bibinfo {author} {\bibfnamefont {J.~C.}\ \bibnamefont
			{Slonczewski}},\ }\href@noop {} {\bibfield  {journal} {\bibinfo  {journal}
			{Int. J. Magn}\ }\textbf {\bibinfo {volume} {2}},\ \bibinfo {pages} {85}
		(\bibinfo {year} {1972})}\BibitemShut {NoStop}%
	\bibitem [{\citenamefont {Malozemoff}\ and\ \citenamefont
		{Slonzewski}(1979)}]{Malozemoff79}%
	\BibitemOpen
	\bibfield  {author} {\bibinfo {author} {\bibfnamefont {A.~P.}\ \bibnamefont
			{Malozemoff}}\ and\ \bibinfo {author} {\bibfnamefont {J.~C.}\ \bibnamefont
			{Slonzewski}},\ }\href@noop {} {\emph {\bibinfo {title} {Magnetic domain
				walls in bubble materials}}}\ (\bibinfo  {publisher} {Academic Press},\
	\bibinfo {address} {New York},\ \bibinfo {year} {1979})\BibitemShut {NoStop}%
	\bibitem [{Note3()}]{Note3}%
	\BibitemOpen
	\bibinfo {note} {The velocity of DW is calculated as $\protect \overline
		{V}=T^{-1}\DOTSI \intop \ilimits@ _{0}^{T}\protect \mathaccentV
		{dot}05F{q}(\tau )\protect \mathrm {d}\tau $, where $\protect \mathaccentV
		{dot}05F{q}(\tau )$ is extracted from numerical simulations and $T\gg 1$ is a
		time simulation.}\BibitemShut {Stop}%
	\bibitem [{\citenamefont {Kravchuk}(2014)}]{Kravchuk14}%
	\BibitemOpen
	\bibfield  {author} {\bibinfo {author} {\bibfnamefont {V.~P.}\ \bibnamefont
			{Kravchuk}},\ }\href {\doibase 10.1016/j.jmmm.2014.04.073} {\bibfield
		{journal} {\bibinfo  {journal} {Journal of Magnetism and Magnetic Materials}\
		}\textbf {\bibinfo {volume} {367}},\ \bibinfo {pages} {9} (\bibinfo {year}
		{2014})}\BibitemShut {NoStop}%
	\bibitem [{\citenamefont {D{\"o}ring}(1948)}]{Doering48}%
	\BibitemOpen
	\bibfield  {author} {\bibinfo {author} {\bibfnamefont {W.}~\bibnamefont
			{D{\"o}ring}},\ }\href {http://www.znaturforsch.com/aa/v03a/c03a.htm}
	{\bibfield  {journal} {\bibinfo  {journal} {Zeitschrift f{\"u}r
				Naturforschung}\ }\textbf {\bibinfo {volume} {3A}},\ \bibinfo {pages} {373}
		(\bibinfo {year} {1948})}\BibitemShut {NoStop}%
	\bibitem [{\citenamefont {Yershov}\ \emph {et~al.}(2016)\citenamefont
		{Yershov}, \citenamefont {Kravchuk}, \citenamefont {Sheka},\ and\
		\citenamefont {Gaididei}}]{Yershov16}%
	\BibitemOpen
	\bibfield  {author} {\bibinfo {author} {\bibfnamefont {K.~V.}\ \bibnamefont
			{Yershov}}, \bibinfo {author} {\bibfnamefont {V.~P.}\ \bibnamefont
			{Kravchuk}}, \bibinfo {author} {\bibfnamefont {D.~D.}\ \bibnamefont {Sheka}},
		\ and\ \bibinfo {author} {\bibfnamefont {Y.}~\bibnamefont {Gaididei}},\
	}\href {\doibase 10.1103/PhysRevB.93.094418} {\bibfield  {journal} {\bibinfo
			{journal} {Physical Review B}\ }\textbf {\bibinfo {volume} {93}},\ \bibinfo
		{pages} {094418} (\bibinfo {year} {2016})}\BibitemShut {NoStop}%
\end{thebibliography}

%



\section*{Supplementary material}

# Supplementary to
# "Domain wall diode based on functionally graded Dzyaloshinskii–Moriya interaction"

Kostiantyn V. Yershov,[1, 2, a)] Volodymyr P. Kravchuk,[1, 3, b)] Denis D. Sheka,[4, c)] Jeroen van den Brink,[2, 5, d)] and Avadh Saxena[6, e)]

[1)] *Bogolyubov Institute for Theoretical Physics of National Academy of Sciences of Ukraine, 03143 Kyiv, Ukraine*
[2)] *Leibniz-Institut für Festkörper- und Werkstoffforschung, IFW Dresden, D-01171 Dresden, Germany*
[3)] *Institut für Theoretische Festkörperphysik, Karlsruher Institut für Technologie, D-76131 Karlsruhe, Germany*
[4)] *Taras Shevchenko National University of Kyiv, 01601 Kyiv, Ukraine*
[5)] *Institute for Theoretical Physics, TU Dresden, 01069 Dresden, Germany*
[6)] *Theoretical Division, Los Alamos National Laboratory, Los Alamos, NM 87545, USA*

The supplementary information provides details on analytical calculations and numerical simulations.

## I. STATIC MAGNETIZATION DISTRIBUTION

Static magnetization distribution is determined by the equations $\delta\mathcal{E}/\delta\theta = 0$ and $\delta\mathcal{E}/\delta\phi = 0$. Taking into account the form of energy density (2) one obtains the following set of equations:

$$\theta'' - \sin\theta\cos\theta\left[(\phi')^2 + 1 + \varepsilon\sin^2\phi\right] \\ + \frac{d'}{2}\cos\phi - d\phi'\sin^2\theta\sin\phi = 0, \quad \text{(S1a)}$$

$$\phi'' + 2\theta'(\cot\theta\phi' + d\sin\phi) - \varepsilon\sin\phi\cos\phi \\ + \frac{d'}{2}\cot\theta\sin\phi = 0. \quad \text{(S1b)}$$

Equation (S1b) has solutions $\phi = \phi_0 = \{0, \pi\}$. Substitution of this solution into (S1a) results in Eq. (4), which determines the function $\theta(\xi)$.

## II. DETAILS OF THE $q$–$\Phi$ MODEL

Magnetization dynamics is studied by means of the phenomenological Landau–Lifshitz–Gilbert equations with additional Zhang-Li spin-transfer torque terms [1–3]

$$-\sin\theta\left(\dot\theta + \tilde{u}\theta'\right) = \frac{\delta\mathcal{E}}{\delta\phi} + \left(\alpha\dot\phi + \tilde{u}\beta\phi'\right)\sin^2\theta, \\ \sin\theta\left(\dot\phi + \tilde{u}\phi'\right) = \frac{\delta\mathcal{E}}{\delta\theta} + \alpha\dot\theta + \tilde{u}\beta\theta'. \quad \text{(S2)}$$

The equations of motion (S2) are the Lagrange–Rayleigh equations

$$\frac{\partial\mathcal{L}}{\partial X_i} - \frac{d}{dt}\frac{\partial\mathcal{L}}{\partial \dot{X}_i} = \frac{\partial\mathcal{F}}{\partial \dot{X}_i}, \qquad X_i \in \{\theta, \phi\} \quad \text{(S3)}$$

for the Lagrange function [4, 5]

$$\mathcal{L} = -\int_{-\infty}^{\infty} \phi\sin\theta\left(\dot\theta + \tilde{u}\theta'\right)d\xi - \frac{\mathcal{E}}{\ell\mathcal{S}} \quad \text{(S4)}$$

and the dissipation function [6]

$$\mathcal{F} = \frac{\alpha}{2}\int_{-\infty}^{\infty}\left[\dot\theta^2 + \sin^2\theta\dot\phi^2\right]d\xi \\ + \tilde{u}\beta\int_{-\infty}^{\infty}\left[\theta'\dot\theta + \sin^2\theta\phi'\dot\phi\right]d\xi. \quad \text{(S5)}$$

Substituting the Ansatz (5) into (S4) and (S5) and performing the integration over $\xi$ we obtain the Lagrange and dissipation functions in the form

$$\mathcal{L} = 2p\Phi\left(\dot{q} - \tilde{u}\right) - \frac{\mathcal{E}}{\ell\mathcal{S}}, \\ \mathcal{F} = \frac{\alpha}{\Delta}\left[\dot{q}^2 + \left(\Delta\dot\Phi\right)^2 + c\dot\Delta^2\right] - 2\tilde{u}\beta\frac{\dot{q}}{\Delta}, \quad \text{(S6)}$$

where $c = \pi^2/12$. Substituting (S6) into the Lagrange–Rayleigh equations (S3) results in the equations of motion

$$\frac{\alpha}{\Delta}\dot{q} + p\dot\Phi = -p\frac{\pi}{2}\frac{\partial d(q)}{\partial q}\cos\Phi + \frac{\beta}{\Delta}\tilde{u}, \\ p\dot{q} - \alpha\Delta\dot\Phi = -p\frac{\pi}{2}d(q)\sin\Phi + \varepsilon\Delta\sin 2\Phi + p\tilde{u}, \quad \text{(S7)} \\ 2c\,\alpha\,\dot\Delta = \frac{1}{\Delta} - \Delta\left(1 + \varepsilon\sin^2\Phi\right).$$

The third equation in (S7) shows that $\Delta$ relaxes towards its equilibrium value $\Delta = 1/\sqrt{1 + \varepsilon\sin^2\Phi}$. The characteristic time $\tau_\text{relax}$ of this relaxation is proportional to damping, $\tau_\text{relax} \propto \alpha$ [7]. Usually $\alpha \ll 1$, therefore one can conclude that the DW width is a slave variable $\Delta(\tau) = \Delta[\Phi(\tau)]$. Typical behavior of DWs dynamics is presented in Fig. S1.

---

a) Corresponding author: yershov@bitp.kiev.ua
b) vkravchuk@bitp.kiev.ua
c) sheka@knu.ua
d) j.van.den.brink@ifw-dresden.de
e) avadh@lanl.gov

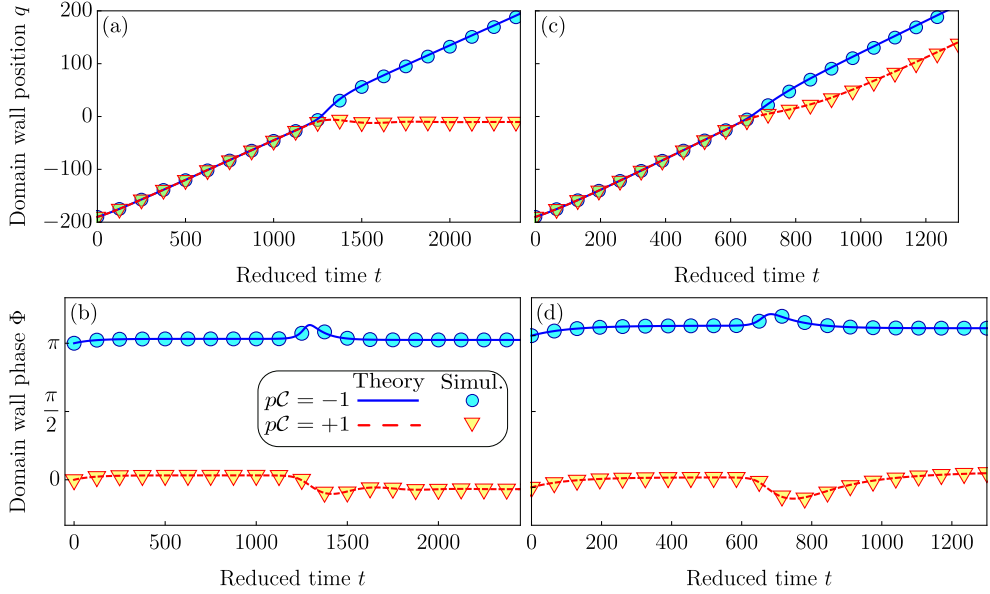

FIG. S1. (Color online) (a),(b) "Forbidden" regime, i.e. DW with $p\mathcal{C} = +1$ is pinned; (c),(d) all DWs are moving. Lines correspond to the solutions of Eqs. (7) and symbols to the results of numerical simulations. In all simulations we use $\alpha = \beta = 0.02$, $d_0 = 0.1$, $w = 10$, $p = +1$, $\varepsilon = 0.25$, and stripes of width $2\ell$. The current is as follows: (a),(b) $\tilde{u} = 0.1$; (c),(d) $\tilde{u} = 0.2$. The corresponding dynamics is illustrated in the Supplemental Video.

### A. Calculation of pinning position and phase

In the limiting case of constant DMI $d' = 0$, static Eqs. (S1) have a solution of a DW with the phase $\cos \Phi_0 = \mathcal{C} = \pm 1$. Therefore, to analyze the dynamics of DW in stripes with a non-zero gradient of the DMI we will use small-angle approximation $\varphi(\tau) = \Phi(\tau) - \Phi_0$ which is valid for stripes with biaxial anisotropy. Using this approximation, equations of motion (7) can be written in the form

$$\alpha \dot{q} + p\dot{\varphi} = -p\mathcal{C}\frac{\pi}{2}\frac{\partial d(q)}{\partial q} + \beta \tilde{u},$$
$$p\dot{q} - \alpha \dot{\varphi} = -p\mathcal{C}\frac{\pi}{2}d(q)\varphi + 2\varepsilon\,\varphi + p\tilde{u}. \quad (S8)$$

For the case of DW pinning, left side of Eqs. (S8) is equal to zero. Therefore, the DW pinning position and phase are defined as

$$d(q_{\text{pin}}) - d(q_0) \approx p\mathcal{C}\frac{2}{\pi}\beta \tilde{u}\,(q_{\text{pin}} - q_0),$$
$$\Phi_{\text{pin}} - \Phi_0 \approx p\frac{\tilde{u}}{p\mathcal{C}\frac{\pi}{2}d(q_0) + \beta\tilde{u}\,(q_{\text{pin}} - q_0) - 2\varepsilon}, \quad (S9)$$

where $q_0$ is the initial DW position. The domain wall pinning position $q_{\text{pin}}$ and phase $\Phi_{\text{pin}}$ as functions of applied current are presented in Fig. S2.

### B. Calculation of eigenfrequencies and effective friction

The equations of motion linearized in the vicinity of the pinning position $q(\tau) = q_{\text{pin}} + \tilde{q}(\tau)$ and phase $\Phi(\tau) = \Phi_{\text{pin}} + \tilde{\Phi}(t)$ read as

$$(1+\alpha^2)\left\|\begin{matrix}\dot{\tilde{q}}\\ \dot{\tilde{\Phi}}\end{matrix}\right\| = \hat{\mathcal{M}}\left\|\begin{matrix}\tilde{q}\\ \tilde{\Phi}\end{matrix}\right\|,$$
$$\hat{\mathcal{M}} = \frac{\pi}{2}\left\|\begin{matrix}-p\alpha d''(q_{\text{pin}})\cos\Phi_{\text{pin}} - d'(q_{\text{pin}})\sin\Phi_{\text{pin}} & p\alpha d''(q_{\text{pin}})\sin\Phi_{\text{pin}} - d(q_{\text{pin}})\cos\Phi_{\text{pin}} + p\dfrac{4}{\pi}\varepsilon\cos 2\Phi_{\text{pin}}\\ p\alpha d'(q_{\text{pin}})\sin\Phi_{\text{pin}} - d''(q_{\text{pin}})\cos\Phi_{\text{pin}} & p\alpha d(q_{\text{pin}})\cos\Phi_{\text{pin}} + d''(q_{\text{pin}})\sin\Phi_{\text{pin}} - \alpha\dfrac{4}{\pi}\varepsilon\cos 2\Phi_{\text{pin}}\end{matrix}\right\|. \quad (S10)$$

The solution of (S10) results in harmonic decaying oscillations with frequency

$$\Omega = \frac{\sqrt{\left|4\operatorname{Det}\left(\hat{\mathcal{M}}\right) - \operatorname{Tr}^2\left(\hat{\mathcal{M}}\right)\right|}}{2(1+\alpha^2)}, \quad (S11)$$

and effective friction

$$\eta = -\frac{\operatorname{Tr}\left(\hat{\mathcal{M}}\right)}{2(1+\alpha^2)}. \quad (S12)$$

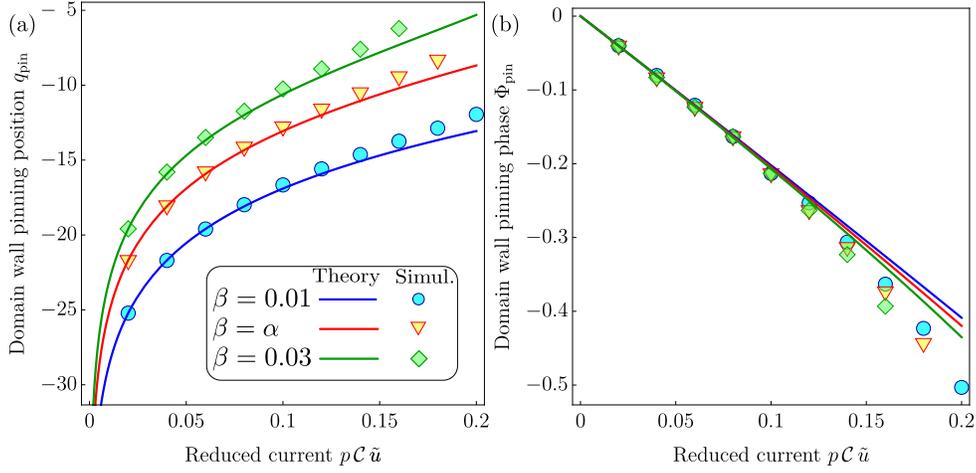

FIG. S2. (Color online) (a) Position of DW pinning as a function of current and (b) phase of DW pinning as a function of current. Solid lines correspond to predictions (S9). Symbols show the results of numerical simulations. In all simulations we use $\varepsilon = 0.25$, $\alpha = 0.02$, $p = +1$, $\tilde{u} > 0$, $d_0 = 0.1$, $w = 10$, and stripes with width $2\ell$.

For the case of low damping $\alpha$, one can calculate the frequency of DW oscillations as

$$\Omega \approx \frac{\pi}{2}\sqrt{d''(q_{\text{pin}})\left[p\mathcal{C}\varepsilon\frac{4}{\pi} - d\left(q_{\text{pin}}\right)\right]}. \quad (S13)$$

### C. DW dynamics for the case $d = \text{const}$

For the case $d = d_0$ equations of motion (7) can be written as

$$\begin{aligned}\frac{\alpha}{\Delta}\dot{q} + p\dot{\Phi} &= \frac{\beta}{\Delta}\tilde{u}, \\ p\dot{q} - \alpha\Delta\dot{\Phi} &= -p\frac{\pi}{2}d_0\sin\Phi + \varepsilon\Delta\sin 2\Phi + p\tilde{u}.\end{aligned} \quad (S14)$$

This set of equations has a traveling wave regime with $\dot{q} = V$ and $\dot{\Phi} = 0$. The corresponding DW velocity is defined as

$$V = \frac{\beta}{\alpha}\tilde{u}, \quad (S15a)$$

while the phase can be found as a solution of the equations

$$\frac{\beta - \alpha}{\alpha}\tilde{u} = \mathcal{G}(\Phi), \quad \mathcal{G}(\Phi) = -\frac{\pi}{2}d_0\sin\Phi + p\varepsilon\Delta\sin 2\Phi. \quad (S15b)$$

Using small-angle approximation $\Phi = \Phi_0 + \varphi$, one can solve Eq. (S15b) with

$$\varphi = \frac{2p}{4\varepsilon - p\mathcal{C}\pi d_0}\left(\frac{\beta - \alpha}{\alpha}\right)\tilde{u}. \quad (S15c)$$

The traveling wave solution (S15) exists for the current less than the Walker current $\tilde{u}_{\text{w}}$ [3, 8]. For $\tilde{u} = \tilde{u}_{\text{w}}$ the function $\mathcal{G}$ reaches its maximum for some $\Phi = \Phi_{\max}$. The maximum phase $\Phi_{\max}$ should be found from the equation $\partial\mathcal{G}/\partial\Phi = 0$. For the DMI free case, the solution of this equation results in the maximal value of phase

$$\cos\Phi_{\max} = \frac{\varepsilon + 1 - \sqrt{\varepsilon + 1}}{\varepsilon}, \quad (S16)$$

which results in the Walker current

$$\tilde{u}_{\text{w}}^0 = 2\frac{\alpha}{|\alpha - \beta|}\sqrt{2 + \varepsilon - 2\sqrt{1 + \varepsilon}} \approx \frac{\alpha}{|\alpha - \beta|}\varepsilon. \quad (S17)$$

For the non-zero value of the DMI strength the Walker current (S17) is modified in the following way (first order perturbation theory for $d_0 \ll \varepsilon$)

$$\tilde{u}_{\text{w}} = \begin{cases} \tilde{u}_{\text{w}}^0 + d_0\dfrac{\alpha}{|\alpha - \beta|}\dfrac{\pi}{2}\dfrac{1}{\sqrt{1 + \sqrt{1 + \varepsilon}}}, & \text{favorable DW}, \\ \tilde{u}_{\text{w}}^0 - d_0\dfrac{\alpha}{|\alpha - \beta|}\dfrac{\pi}{2}\dfrac{1 + \varepsilon - \sqrt{1 + \varepsilon}}{\sqrt{\varepsilon(1 + \varepsilon)\left(\sqrt{1 + \varepsilon} - 1\right)}}, & \text{unfavorable DW}. \end{cases} \quad (S18)$$

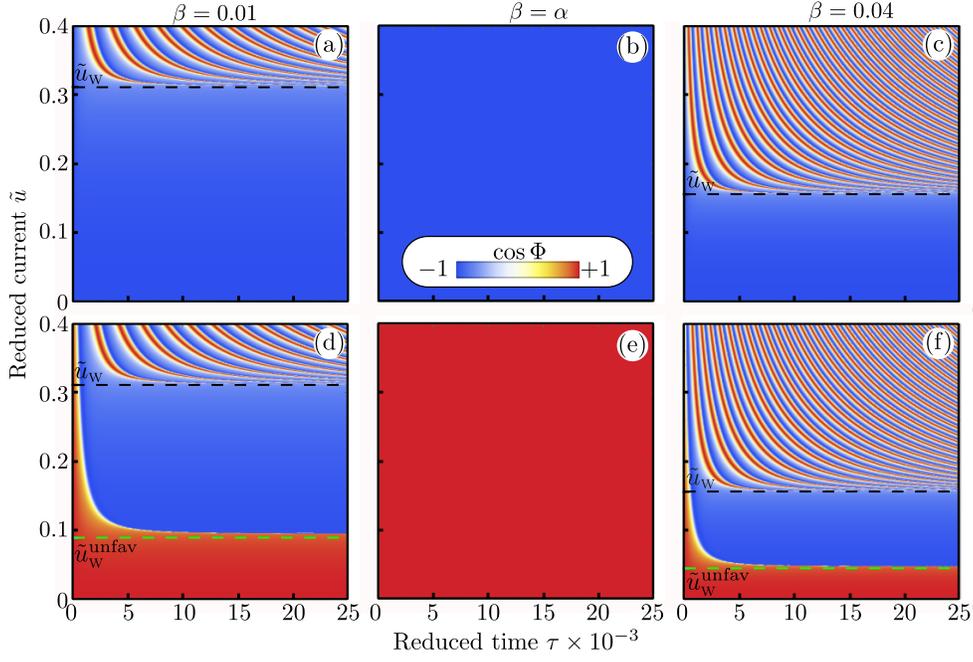

FIG. S3. (Color online) Evolution of $\cos\Phi$ for DW for different values of current in terms of density plot for favorable (top row) and unfavorable (bottom row) DWs [data obtained from solutions of the collective variables equations (S14)]. Horizontal dashed lines correspond to the analytical prediction (S19). Parameters of the calculations are as follows: $\alpha = 0.02$, $d_0 = 0.05$, $p = +1$, $\varepsilon = 0.1$, and stripe width $2\ell$.

In the limiting case of small easy-plane anisotropy we have

$$\tilde{u}_{\text{W}} \approx \frac{\alpha}{|\alpha - \beta|}\left(\varepsilon \pm \frac{\pi}{2\sqrt{2}}d_0\right), \quad \text{(S19)}$$

where "+" sign corresponds to the favorable DW, while "−" sign corresponds to the unfavorable DW. The value of the Walker current for unfavorable DW is smaller as compared to the favorable one and it defines the current of the DW phase flip, i.e. $\tilde{u}_{\text{W}}^{\text{unfav}} < \tilde{u} < \tilde{u}_{\text{W}}$ and we have traveling wave motion with a single flip of the phase with $\Phi \to \Phi + \pi$, see Fig. S3. It is necessary to mention that for the case $\beta = \alpha$ both DWs move in a traveling-wave regime without any flip of the phase $\Phi$, i.e. $\tilde{u}_{\text{W}}(\beta = \alpha) \to \infty$, see Fig. S3(b) and S3(e).

Typical behavior of DW dynamics under the action of spin-polarized current for the cases with $\tilde{u} < \tilde{u}_{\text{W}}$ and $\tilde{u} > \tilde{u}_{\text{W}}$ is presented in Fig. S4.

## III. DETAILS OF NUMERICAL SIMULATIONS

In order to verify our analytical calculations we perform a set of numerical simulations. We consider a ribbon with square lattice and lattice constant $a$. Each node is characterized by a magnetic moment $\boldsymbol{m_n}(t)$ which is located at position $\boldsymbol{r_n}$. Here $\boldsymbol{n} = (i, j)$ is a two dimensional vector which defines the magnetic moment and its position on a lattice with size $N_1 \times N_2$ ($i \in [1, N_1]$ and $j \in [1, N_2]$). Magnetic moments are ferromagnetically coupled. The dynamics of magnetic system is governed by the Landau–Lifshitz–Gilbert equation with additional Zhang–Li spin-torque terms [1–3]

$$\begin{aligned}\frac{\mathrm{d}\boldsymbol{m_n}}{\mathrm{d}\tau} =& \boldsymbol{m_n} \times \frac{\partial\mathscr{H}}{\partial\boldsymbol{m_n}} + \alpha\,\boldsymbol{m_n} \times \frac{\mathrm{d}\boldsymbol{m_n}}{\mathrm{d}\tau}\\ &+ \tilde{u}\,\boldsymbol{m_n}\times\left[\boldsymbol{m_n}\times\boldsymbol{m_{q+(i+1,j)}}\right] + \beta\tilde{u}\,\boldsymbol{m_n}\times\boldsymbol{m_{n+(i+1,j)}},\end{aligned} \quad \text{(S20)}$$

where $\tau = \omega_0 t$ is a reduced time with $\omega_0 = |\gamma_0|K/M_s$, $\tilde{u} = jP\mu_{\text{B}}/(|e|M_s\omega_0\ell)$ is a dimensionless current, $\alpha$ and $\beta$ denote Gilbert damping and the nonadiabatic spin-transfer parameter, respectively. Here $\mathscr{H}$ is a dimensionless energy normalized by $K$. We consider three contributions to the energy of the system

$$\mathscr{H} = \mathscr{H}_{\text{EX}} + \mathscr{H}_{\text{A}} + \mathscr{H}_{\text{D}}. \quad \text{(S21a)}$$

The first term in (S21a) is the exchange energy

$$\mathscr{H}_{\text{EX}} = -\frac{1}{2}\frac{\ell^2}{a^2}\sum_{\boldsymbol{n}}\boldsymbol{m_n}\cdot\boldsymbol{m_{n+\delta}}, \quad \text{(S21b)}$$

where $\boldsymbol{\delta}$ runs over nearest neighbors.

The second term in (S21a) is the anisotropy energy

$$\mathscr{H}_{\text{A}} = -\sum_{\boldsymbol{n}}\left[(\boldsymbol{m_n}\cdot\hat{\boldsymbol{z}})^2 - \varepsilon(\boldsymbol{m_n}\cdot\hat{\boldsymbol{y}})^2\right], \quad \text{(S21c)}$$

where $\varepsilon$ is the anisotropy ratio.

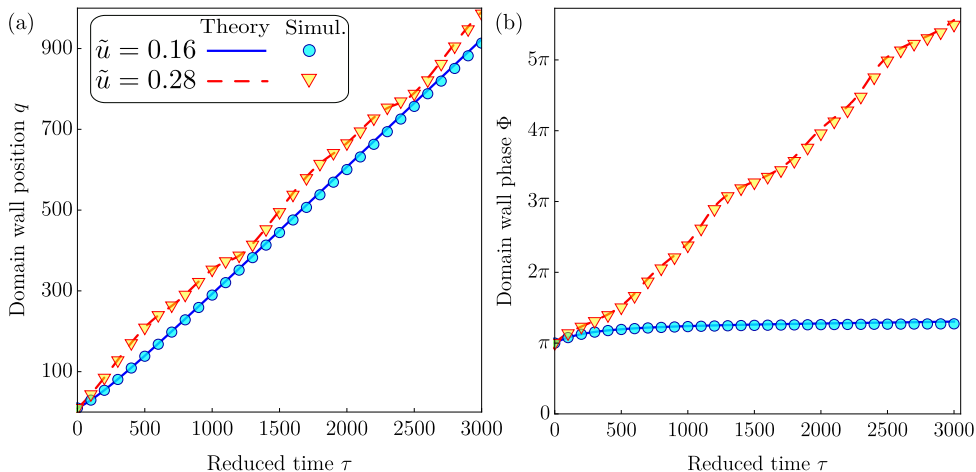

FIG. S4. (Color online) Different regimes of the domain wall motion in terms of the collective variables. Parameters are the same as in Fig. 3 [main text] and $\beta = 0.04$. Solid and dashed lines correspond to the traveling-wave ($\tilde{u} = 0.16 < \tilde{u}_\text{w}$) and precession ($\tilde{u} = 0.28 > \tilde{u}_\text{w}$) regimes, respectively. Symbols and lines correspond to simulation data and solutions of Eqs. (S14), respectively.

The last term in Eq. (S21a) is the DMI energy

$$\mathscr{H}_\text{D} = \frac{1}{2} \frac{\ell}{a} \sum_{\boldsymbol{n}} d\left(\boldsymbol{n}\right) \boldsymbol{d} \cdot \left[\boldsymbol{m_n} \times \boldsymbol{m_{n+\delta}}\right], \qquad (\text{S21d})$$

with $d\left(\boldsymbol{n}\right) = D\left(\boldsymbol{n}\right)/\sqrt{AK}$, and $\boldsymbol{d} = \hat{\boldsymbol{z}} \times (\boldsymbol{\delta} - \boldsymbol{n})/a$ being the DMI vector.

The dynamical problem is considered as a set of $3N_1 N_2$ ordinary differential equations (S20) with respect to $3N_1 N_2$ unknown functions $m^\text{x}_{\boldsymbol{q}}(t)$, $m^\text{y}_{\boldsymbol{q}}(t), m^\text{z}_{\boldsymbol{q}}(t)$. For given initial conditions, the set (S20) is integrated numerically. During the integration process, the condition $|\boldsymbol{m_q}(t)| = 1$ is controlled. We considered a stripe with $N_1 = 5001$, $N_2 = 21$, and magnetic length $\ell = 10a$.

We consider the DMI parameter profile in the form

$$d\left(i\right) = \frac{d_0}{2} \left[\tanh\left(\frac{i}{w}\right) + 1\right], \qquad (\text{S22})$$

where $d_0$ is the amplitude of DMI strength and $w$ is a width.

The numerical experiment consists of two steps. Initially, we relax the DW structure in an over-damped regime ($\alpha = 0.5$) in the area with $d \approx 0$ and $\tilde{u} = 0$.

In the second step, we simulate current induced dynamics with Gilbert damping $\alpha = 0.02$ for different values of nonadiabatic spin-transfer parameter $\beta$, current $\tilde{u}$, and DMI amplitude. The DW position $q$ is determined as a fitting parameter for the function $m_z(i) = -p\tanh\left[(i-q)/\Delta\right]$, then the phase is determined from the equation $\tan\Phi = m_x(q)/m_y(q)$.

### A. Movies of the DW motion

For a better illustration of the influence of functionally graded profile of DMI strength we prepared movies of DW dynamics. Data in movies correspond to the results obtained by means of numerical simulations. Here in all simulations we used the profile of DMI strength defined in (S22) with $d_0 = 0.1$ and $w = 10$. Other parameters are $\beta = 0.03$ and $\alpha = 0.02$. In all cases DWs were in the position $q$ with $d(q) \approx 0$ (far from the place with $d' \neq 0$).

- `movie_u_0.1_beta_0.03_pinning.mp4` shows dynamics of two DWs in a ferromagnetic stripe. Here we have DWs with $p\mathcal{C} = +1$ (*unfavorable*) and $p\mathcal{C} = -1$ (*favorable*). The applied current is $\tilde{u} = 0.1$. DW with $p\mathcal{C} = +1$ is pinned in the stripe, while DW with $p\mathcal{C} = -1$ moves.

- `movie_u_0.2_beta_0.03_breakdown.mp4` illustrates the current driven DWs motion with $\tilde{u} = 0.2$. Here, both DWs are not pinned, i.e. DW with $p\mathcal{C} = +1$ demonstrates a decrease of velocity in the area with $d' \neq 0$. Afterward it moves with the same velocity as DW with $p\mathcal{C} = -1$.

- `movie_u_0.1_beta_0.03_trap.mp4` shows the trap for DW. Here we used the DMI profile in the form

$$d\left(i\right) = \frac{d_0}{2} \left[\tanh\left(\frac{i}{w}\right) - \tanh\left(\frac{i-\delta}{w}\right)\right],$$

where $\delta \gg \Delta$ defines the width of area with non-zero DMI strength. In the simulations we have DW with $p\mathcal{C} = -1$. Firstly, we applied a current $\tilde{u} = 0.1$ and observed current induced motion of DW in the area with $d = d_0$ and its pinning. Afterward we changed the direction of the current flow ($\tilde{u} = -0.1$) and again DW was pinned. DW is trapped in the area with $|d| > 0$.

[1] Y. B. Bazaliy, B. A. Jones, and S.-C. Zhang, "Modification of the Landau–Lifshitz equation in the presence of a spin–polarized current in colossal- and giant–magnetoresistive materials," Physical Review B **57**, R3213–R3216 (1998).
[2] S. Zhang and Z. Li, "Roles of nonequilibrium conduction electrons on the magnetization dynamics of ferromagnets," Physical Review Letters **93**, 127204 (2004).
[3] A. Thiaville, Y. Nakatani, J. Miltat, and Y. Suzuki, "Micromagnetic understanding of current-driven domain wall motion in patterned nanowires," Europhysics Letters (EPL) **69**, 990–996 (2005).
[4] W. Döring, "Über die trägheit der wände zwischen weißschen bezirken," Zeitschrift für Naturforschung **3A**, 373–379 (1948).
[5] K. V. Yershov, V. P. Kravchuk, D. D. Sheka, and Y. Gaididei, "Curvature and torsion effects in spin-current driven domain wall motion," Physical Review B **93**, 094418 (2016).
[6] T. Gilbert, "A phenomenological theory of damping in ferromagnetic materials," IEEE Transactions on Magnetics **40**, 3443–3449 (2004).
[7] B. Hillebrands and A. Thiaville, eds., *Spin dynamics in confined magnetic structures III*, Topics in Applied Physics, Vol. 101 (Springer, Berlin, 2006).
[8] A. Mougin, M. Cormier, J. P. Adam, P. J. Metaxas, and J. Ferré, "Domain wall mobility, stability and Walker breakdown in magnetic nanowires," EPL (Europhysics Letters) **78**, 57007 (2007).



\end{document}